\documentclass[aps,pre,twocolumn,superscriptaddress,showpacs,floatfix]{revtex4-1}

\usepackage{graphicx}
\usepackage{dcolumn} 
\usepackage{bm}  
\usepackage{latexsym}
\usepackage{amsmath,amstext,amssymb}

\begin{document}

\title{Distributed flow optimization and cascading effects in weighted complex networks}
\author{Andrea Asztalos}
\email{Corresponding author: asztaa@rpi.edu}
\author{Sameet Sreenivasan}
\affiliation{Social and Cognitive Networks Academic Research Center, Rensselaer Polytechnic Institute,\\
110 8$^{th}$ Street, Troy, NY 12180--3590, USA}
\affiliation{Department of Computer Science, Rensselaer Polytechnic Institute, 110 8$^{th}$ Street, Troy, NY 12180--3590, USA}
\affiliation{Department of Physics, Applied Physics and Astronomy, Rensselaer Polytechnic Institute,\\
110 8$^{th}$ Street, Troy, NY 12180--3590, USA}
\author{Boleslaw K. Szymanski}
\affiliation{Social and Cognitive Networks Academic Research Center, Rensselaer Polytechnic Institute,\\
110 8$^{th}$ Street, Troy, NY 12180--3590, USA}
\affiliation{Department of Computer Science, Rensselaer Polytechnic Institute, 110 8$^{th}$ Street, Troy, NY 12180--3590, USA}
\author{G. Korniss}
\affiliation{Social and Cognitive Networks Academic Research Center, Rensselaer Polytechnic Institute,\\
110 8$^{th}$ Street, Troy, NY 12180--3590, USA}
\affiliation{Department of Physics, Applied Physics and Astronomy, Rensselaer Polytechnic Institute,\\
110 8$^{th}$ Street, Troy, NY 12180--3590, USA}

\date{\today}

\begin{abstract}
We investigate the effect of a specific edge weighting scheme $\sim
(k_i k_j)^{\beta}$ on distributed flow efficiency and robustness to
cascading failures in scale-free networks. In
particular, we analyze a simple, yet fundamental distributed flow
model: current flow in random resistor networks. By the tuning of
control parameter $\beta$ and by considering two general cases of
relative node processing capabilities as well as the effect of bandwidth, we show the
dependence of transport efficiency upon the correlations between the
topology and weights. By studying the severity of cascades for
different control parameter $\beta$, we find that network resilience
to cascading overloads and network throughput is optimal for the
same value of $\beta$ over the range of node capacities and available bandwidth.
\end{abstract}

\maketitle

\section{Introduction} 
\label{Sec1}

Distributed flows are ubiquitous in natural and man made systems.
Examples of such distributed flows include but are not limited to: water flow in 
riverbeds, data flow on the Internet, current flow in the power grid, nutrient flows in leaves. 
By studying real-world transport systems, a natural question appears as how such flows can be designed in order
to optimize various transport characteristics, collectively referred
to as objective functions, e.g. searchability \cite{Guimera_PRL02}, transport efficiency
\cite{YanPre06, LopezTransportPRL05}, average packets traveling time
\cite{WangPre06}, resilience against cascading overloads
\cite{YangPRE09} and against damages \cite{Leafprl10}. 

In a large class of transport models in networks, flow is assumed to
be limited to and directed along the shortest paths between the
source and destination
\cite{SameetBottleneck07,DanilaPRE06,YangPRE09}. In another class of
flow models, motivated primarily by search and discovery in networks, routing
of packets can be random, which have led to the studies of regular
\cite{Chandra,Aldous}, weighted
\cite{KornissweightsPRE,HuEPLbandwidth07,LingPhysLettA10,Danila_RW_PRE2006},
or adaptive random walks (RWs) in networks \cite{Asztalos_EPL2010}.
In between these two extremes lies a class of models where transport
and flow is both directed and distributed: packets or information
flow are distributed among {\em all possible paths} emanating from
the source and ending at the target node. In this paper we
systematically study the simplest such example of a directed, distributed flow: {\em currents in
resistor networks}.

Resistor networks are arbitrary networks in which all edges are
resistors with a specific ``electrical" conductance and a pair of
nodes have been designated to be the source and sink (target) of a
current $I$ flowing through the network. They have not only been
used to capture inherent transport capabilities of the underlying
(possibly weighted) complex communication or information networks
\cite{LopezTransportPRL05,Andrade05,DsLee_epl06,KornissSWPhysLettA06,KornissweightsPRE},
but also to describe the transport of carbohydrates in plants
\cite{Plant07}, to investigate flow of fluids in porous media
\cite{PorousFatt56,KirkpatrickPRL71}, to find communities in complex
networks \cite{NewmanGirvan04}, or to construct page-ranking schemes
for search engines \cite{Kaul_ACM2009}.
Further, there are fundamental connections between random walks
(hitting times) and resistor networks (two-point resistance)
\cite{DoyleSnell84,Chandra,Tetali_JTP1991,Redner07,KornissweightsPRE,GBS_SIAM2008,Ellens_LAA2011}.
These fundamental observables (and the corresponding load and
betweenness measures) can also serve as a starting point in routing
schemes in actual communication networks
\cite{TLG_GTC2008,TLG_IEEE2010}. 

Here, we present extensive and
systematic numerical results pertaining to two general cases of
relative node processing capabilities along with the effect of bandwidth on
current flows in weighted resistor networks. In the following section (Sec. \ref{Sec2}) we describe the model we use to study distributed flows. In Sec. \ref{Sec3} we study numerically the effect of two different node capacities and bandwidth upon flow efficiency. In Sec. \ref{Sec4} we present results about maximizing network robustness against cascading failures, and present concluding remarks in Sec. \ref{Sec5}.

\section{Model}
\label{Sec2}
In this paper we consider a specific scheme of assigning
conductances to edges $-$ edge weighting scheme$-$ and study how the
parameter in this scheme affects the efficiency of flow and
robustness of the network to cascading failures. Consider a resistor
network having $N$ nodes and $M$ edges. The conductance of an
arbitrary edge $e=(i,j)$ is set to be proportional to the end-point
degrees, namely $C_{ij} \sim a_{ij} (k_i k_j)^{\beta}$, where
$a_{ij}$ is the $(i,j)^{th}$ element of the network's adjacency
matrix and $\beta$ is a {\em control parameter}, taking only real
values. This choice, also studied before in \cite{Colizza06,
KornissweightsPRE, Baron10}, has been motivated by empirical studies
\cite{Barrat04, WeightNetEivind05} where edge weights
were observed to follow a similar trend. 
An additional motivation is that this scheme provides a convenient way of studying
topologically (structurally) biased flows as a function of
$\beta$ control parameter.  This parameter allows one to bias the
current flow predominantly towards large degree nodes (hubs) (for
positive $\beta$) or to avoid them (for negative $\beta$). When
$\beta=0$, all edge weights are equal, thus the current flow is
solely influenced by the network topology.

Currents along the edges of the network are obtained following the method summarized in \cite{KornissBookCh}.
When $I$ units of current flow into the network at a source $s$ and leave at a target $t$, then for an arbitrary node $i$, charge conservation (Kirchoff's law) combined with Ohm's law dictates:
\begin{equation}
\sum_{j=1}^{N} C_{ij} (V_i -V_j) =I(\delta_{is} -\delta_{it}), \;\; \forall i=1,\dots, N.
\label{eq:charge}
\end{equation}
Keeping in mind that the weighted network Laplacian can be written as $L_{ij}=\delta_{ij} C_i - C_{ij}$, where $C_i= \sum_{j=1}^N a_{ij} C_{ij}$, the system of linear equations (\ref{eq:charge}) can be transformed into the matrix equation
\begin{equation}
L V=\mathcal{I}.
\label{eq:matrixform}
\end{equation}
$V$ is the unknown column voltage vector, while $\mathcal{I}_i$ is the net current flowing into the network at node $i$, which is zero in all cases except the source and target nodes. Equation \ref{eq:matrixform} is solvable for voltages, as long as the inverse of the Laplacian $L$ is known. As the $L$ matrix is singular (all rows and columns of $L$ sum up to zero implying that there is an eigenvalue $\lambda_1=0$ with a corresponding constant eigenvector) it can not be directly inverted. 
This, however presents no technical difficulty, as all relevant physical observables can be expressed in terms of the inverse (or pseudo-inverse) Laplacian, defined in the space orthogonal to the zero mode: $G=L^{-1}$. This is achieved by using spectral decomposition of the network Laplacian \cite{KornissweightsPRE,Slepc_Hernandez}. For example, by choosing the reference potential to be the mean voltage \cite{KornissSWPhysLettA06}, $\hat{V}_i = V_i -\langle V \rangle$, where $\langle V \rangle = (1/N)\sum_{j=1}^N V_j$,  one obtains:
\begin{equation}
\hat{V_i} =(G\mathcal{I})_i =\sum_{j=1}^N G_{ij} I(\delta_{js} -\delta_{jt})= I(G_{is} -G_{it}),
\label{eq:voltage}
\end{equation}
for each node $i$. Thus, for {\em $I$ units of current} and for {\em a given source/target pair}, the current flowing through an arbitrary edge $e=(i,j)$ is 
\begin{equation}
I_{ij}^{st} = C_{ij}  (V_i -V_j) = C_{ij}I (G_{is}-G_{it}-G_{js}+G_{jt}),
\label{eq:current}
\end{equation}
 while the current flowing through node $i$ is given by 
 \begin{equation}
 I_{i}^{st}=\frac{1}{2} \sum_{j} |I_{ij}^{st}|,
 \end{equation}
where the sum is taken over all neighbors of node $i$. So far we have considered the general case of $I$ units of current entering (leaving) the network at a given source (target) pair. For the studies in this paper, with the exception of Sec. \ref{Sec4B}, we assume that {\em unit current} flows between {\em $N$ source/target pairs} simultaneously. Specifically, we assume that all nodes are simultaneously sources and unit current flows into the network at each source. For each source node a target is chosen randomly and uniformly from the remaining $N-1$ nodes. Consequently, the net current flowing through an arbitrary edge/node gives the edge/node current-flow betweenness \cite{Newman_betw, BrandesCflow, KornissBookCh}:
\begin{equation}
\ell_{ij}=\frac{1}{N-1}\sum_{s,t=1}^N |I_{ij}^{st}|, \;\; \ell_i =\frac{1}{N-1}\sum_{s,t=1}^N  |I_i^{st}|.
\end{equation}
Note, $\ell_i =1/2 \sum_j \ell_{ij}$, where the sum is over all neighbors of node $i$.
These quantities capture appropriately the amount of information passing through an edge or a link in a distributed fashion, and therefore we will refer to them as edge and vertex loads. We clarify, this is different from the quantity `vertex load' introduced in \cite{GohLoad01, Goh_PhysA03}, defined as the accumulated number of data packets passing through a vertex when all pairs of nodes send and receive a data packet in unit time transmitted along the shortest paths connecting them.
Currents along the edges and nodes are uniquely determined by the network topology and the weight exponent $\beta$. Therefore, this is a fully
deterministic model and the only source of randomness in the problem is in the network structure.

We restrict our studies to random, uncorrelated scale-free networks constructed using the configuration model \cite{MolloyR95} characterized by a fat-tailed degree distribution $P(k)=ck^{-\gamma}$, where $c$ is the normalization constant and $k_{max} \sim (\langle k \rangle N)^{1/2}$ \cite{BogunaUCM04}.

\section{Vertex and edge load  landscapes}
\label{Sec3}
\begin{figure}[htbp]
\centering
\includegraphics[width=2.5in]{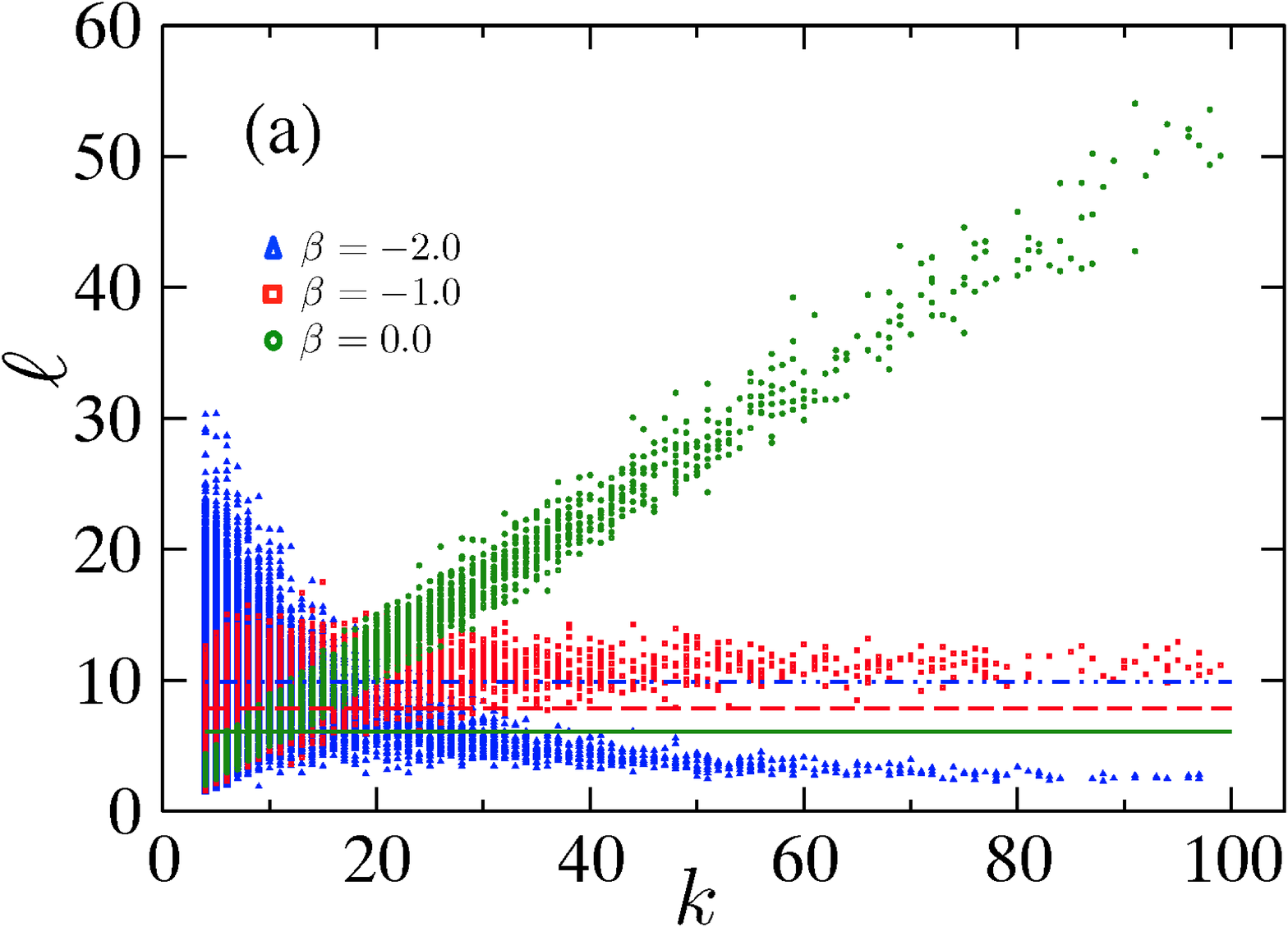}
 \includegraphics[width=2.5in]{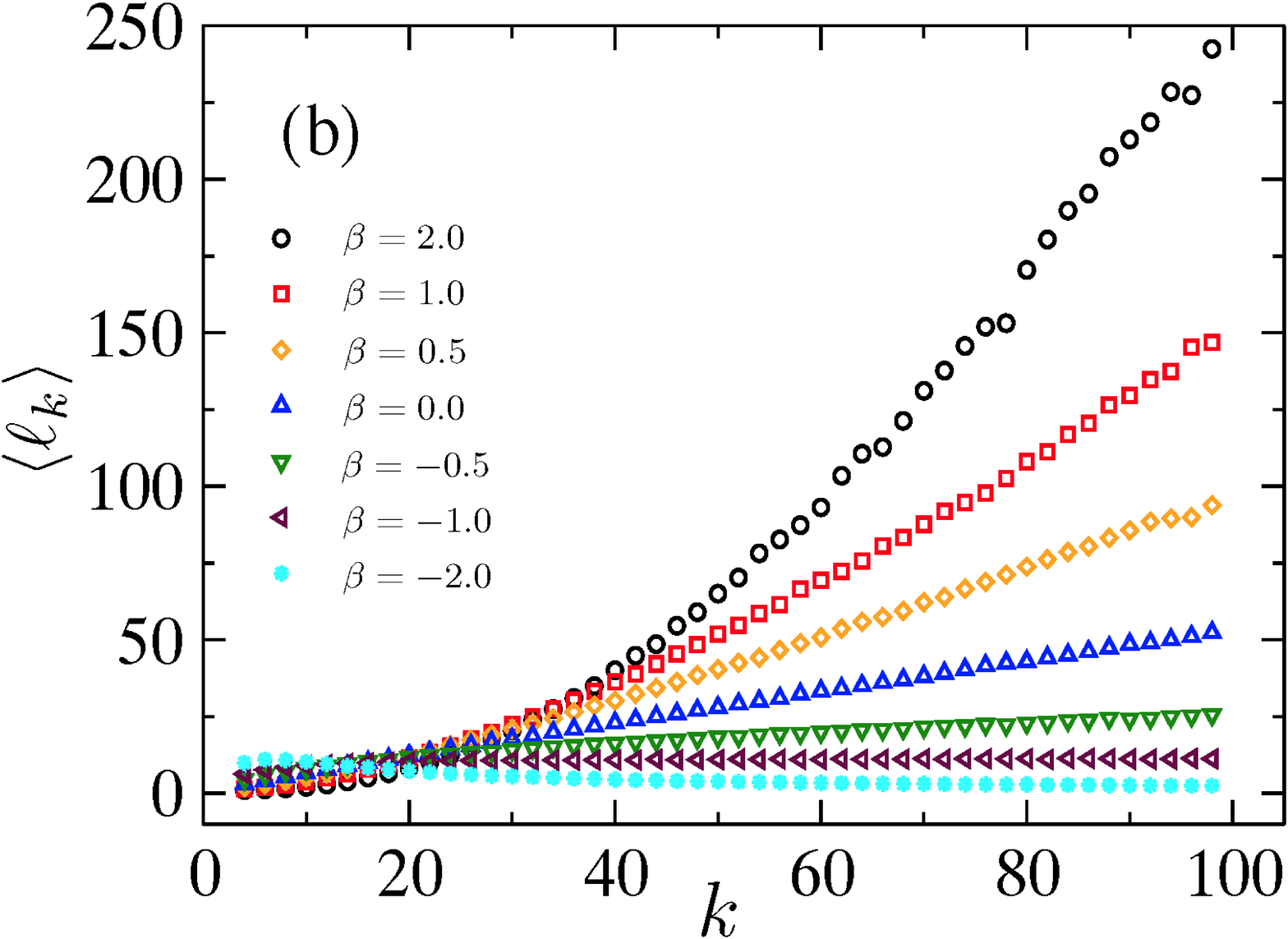}
\includegraphics[width=2.5in]{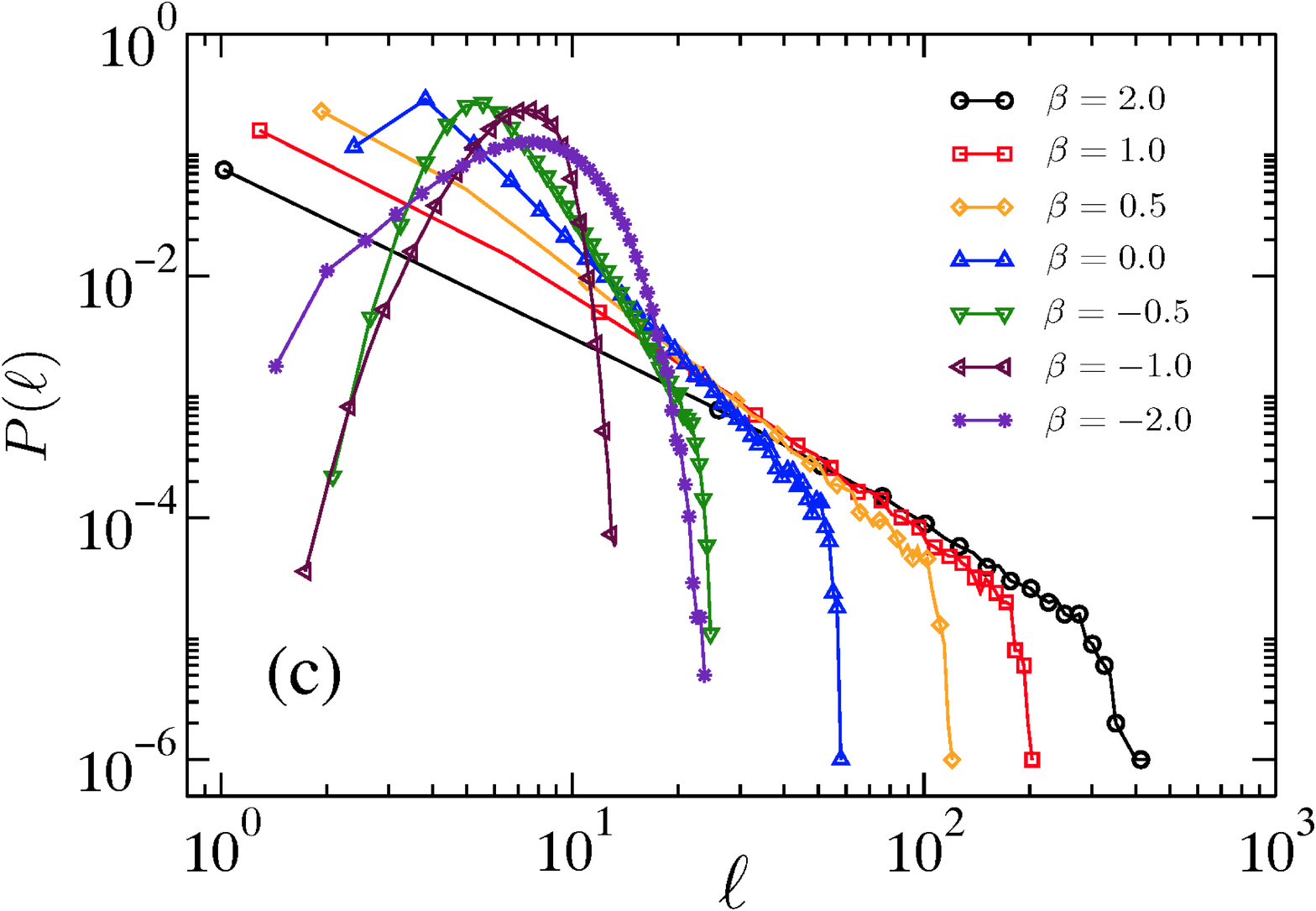}
\includegraphics[width=2.5in]{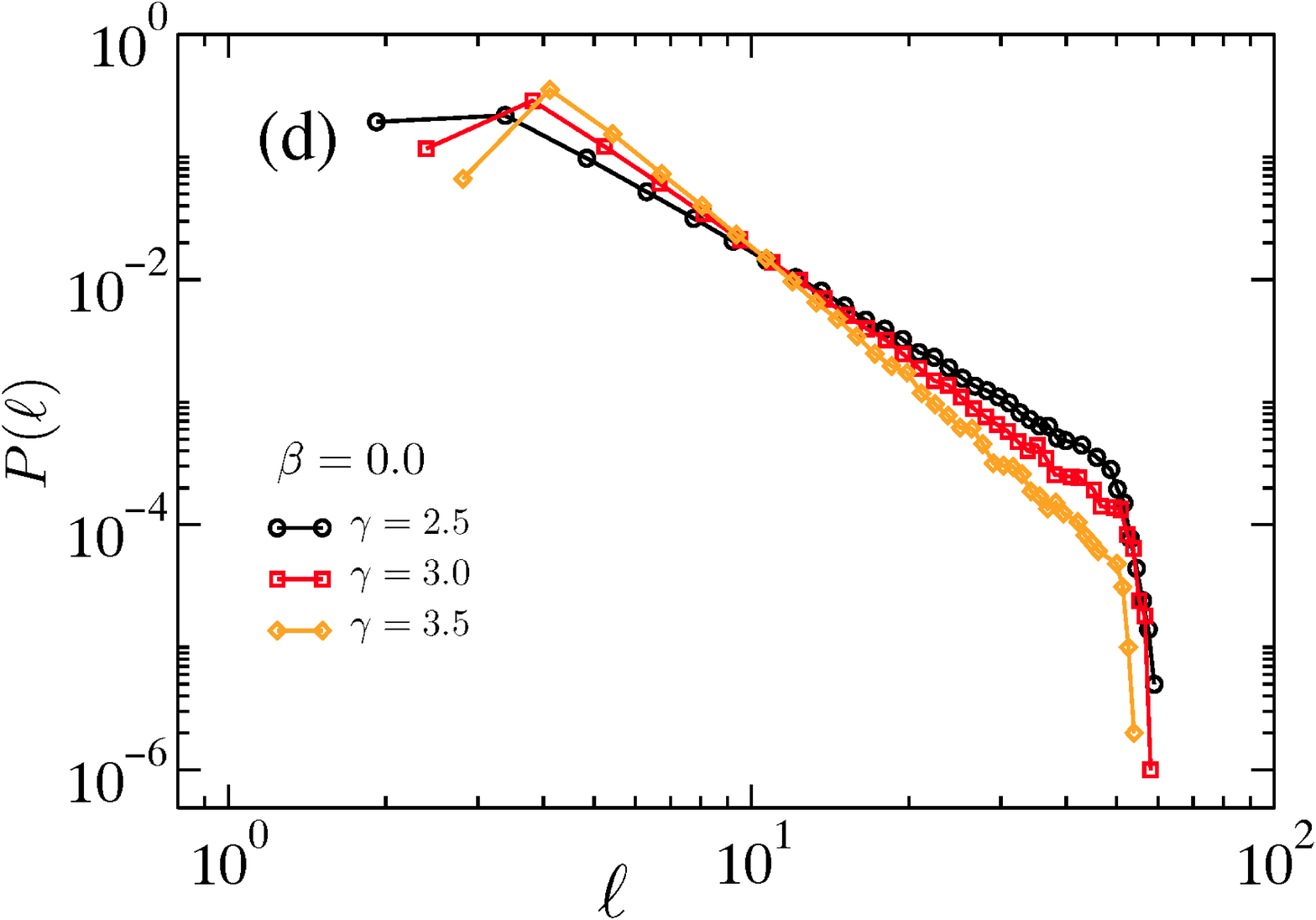}
\caption{ (a) Scatterplot between vertex load and degree showing strong correlation between these variables for unbiased flow $(\beta=0)$. Color coded horizontal lines show the average vertex load in the network: green solid line for $\beta=0.0$, red dashed line for $\beta=-1.0$ and blue dashed line with two dots for $\beta=-2.0$. (b) Average vertex load for different degree classes as we vary $\beta$. Data in both (a) and (b) are plotted for random scale-free networks with $\gamma=2.5$. (c) Vertex-load distributions calculated on scale-free networks with $\gamma=3.0$ for various values of $\beta$. (d) Vertex-load distributions for unbiased flow for different values of degree exponents $\gamma$. Data were averaged over $400$ different network realizations ($N=10^3$, $\langle k \rangle=10$).}
\label{fig1}
\end{figure}

Attributes of the vertex load landscapes for random scale-free networks are shown in Fig. \ref{fig1}. Positive load-degree correlations are observed in Fig. \ref{fig1}(a) for unbiased ($\beta=0$) flow and these correlations become stronger as the flow gets increasingly biased towards the hubs ($\beta>0$) as shown in Fig. \ref{fig1}(b).
The correlations disappear when $\beta \approx -1$; the loads become
balanced, however the average vertex load in the network increases
slightly. As the hub-avoiding bias of flows increases (by
lowering $\beta$) relatively high loads start to appear on small
degree nodes, contributing to an increase in the average vertex
load. The same tendency was previously observed
\cite{Huang_MS2010,KornissBookCh} for Barab\'{a}si-Albert
(BA) scale-free graphs \cite{Science_BA99}. Furthermore, vertex-load
distributions [Fig. \ref{fig1}(c)] show a power-law tail for
unbiased and hub-biased flows with gradually decreasing exponential
cut-offs. The tail significantly diminishes as the flow is biased against hubs and 
towards small degree nodes e.g., for $\beta \approx -0.5$ and disappears for
$\beta \approx -1.$ Note, that for $\beta \approx -1$ the variance of the
distribution is the smallest among all $\beta$ values. A strong
correlation between the degree exponent of the network and the
exponent $\delta$ characterizing the power-law tail of the 
vertex-load distributions for unbiased flow ($\beta=0$) can also be
observed as shown in Fig. \ref{fig1}(d). For $\gamma=2.5$, $3.0$ and
$3.5$ we obtain $\delta= 2.52 \pm 0.01$, $3.01 \pm 0.02$ and $3.59
\pm 0.01$, respectively.

\begin{figure}[htb]
\centerline{\includegraphics[width=2.5in]{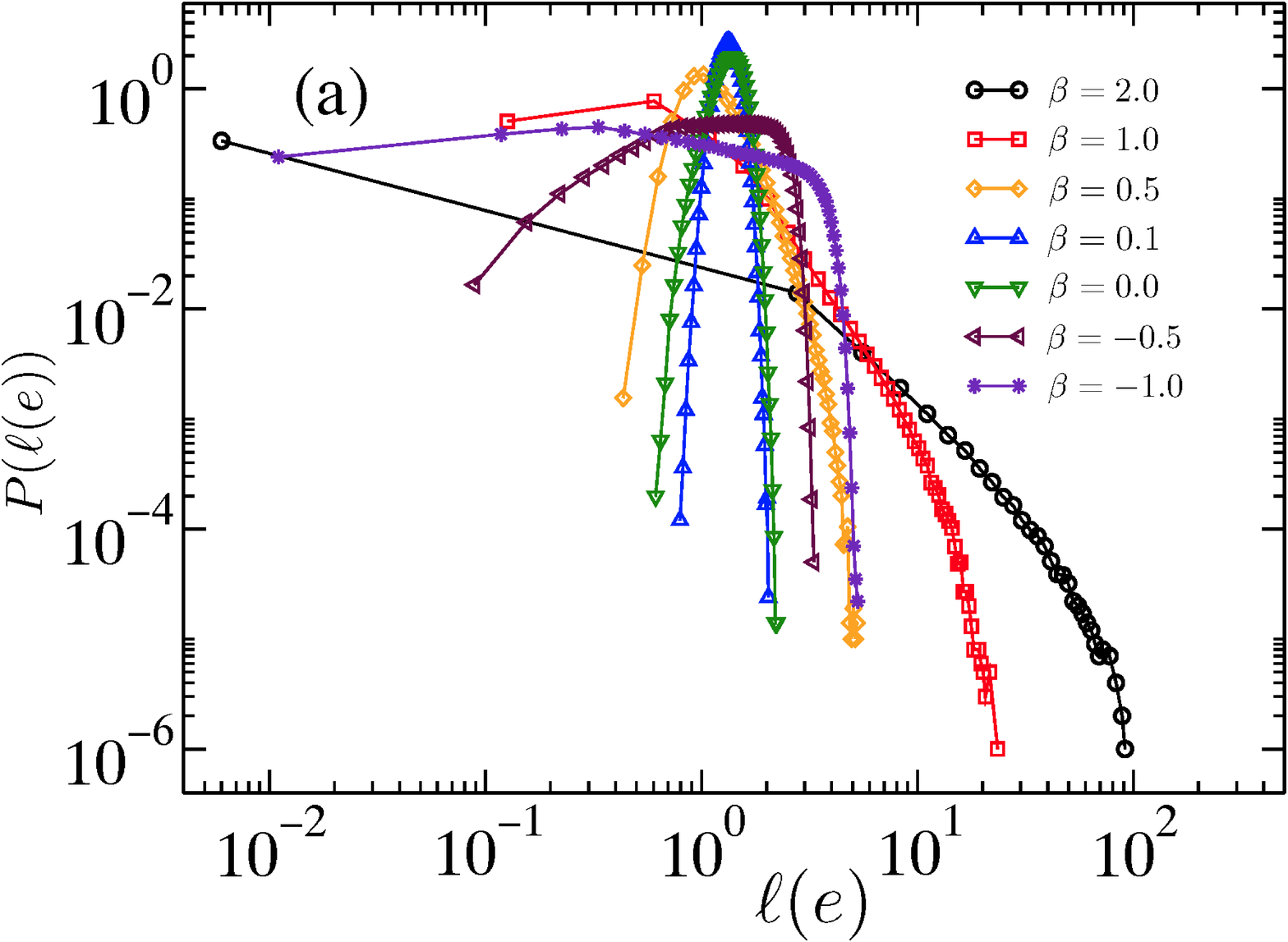}}
\centerline{\includegraphics[width=2.5in]{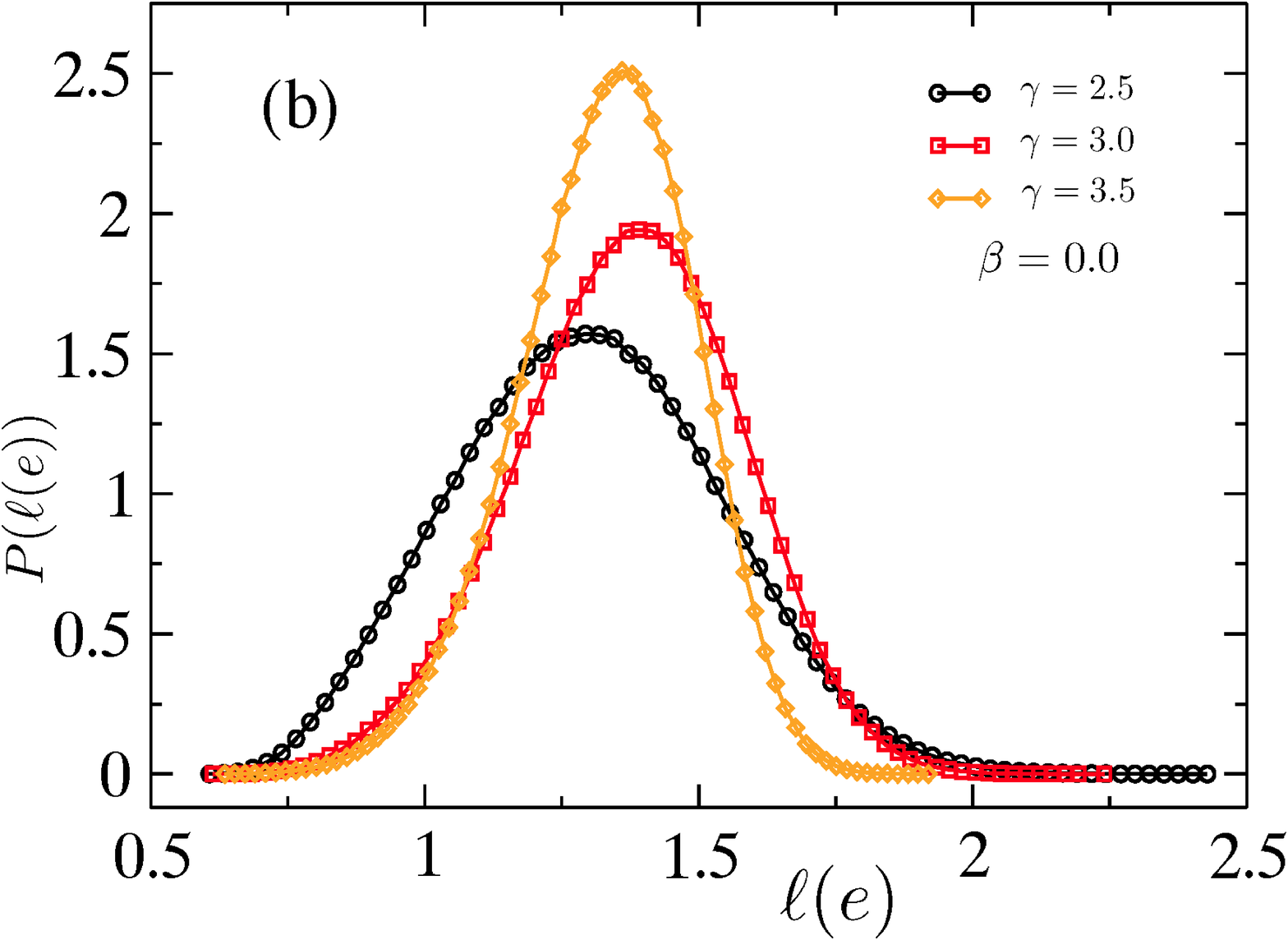}}
\caption{(a) Edge-load distributions in random scale-free networks ($\gamma=3.0$) obtained for various values of $\beta$. (b) Edge-load distributions for unbiased flow ($\beta=0.0$). Network specifications are the same as in Fig. \ref{fig1}.}
\label{fig2}
\end{figure}

The edge-load distributions for various values of $\beta$ values are shown in Fig. \ref{fig2}(a).
Large heterogeneity in the edge-load distribution is observed for strong hub-biased flows ($\beta= 2.0$). This heterogeneity diminishes as the hub bias is decreased and the tail of distribution becomes exponential. In contrast with node loads in Fig. \ref{fig1}(d), the edge loads, for unbiased flows are normally distributed as shown in Fig. \ref{fig2}(b). Interestingly, the edge loads become most balanced for $\beta \approx 0.1$, when flows are slightly biased towards large degree nodes. 

\section{Distributed flow optimization}
\label{Sec4}
\subsection{Optimal network throughput}
\label{Sec4A}
\begin{figure}[htbp]
\centerline{\includegraphics[width=2.5in]{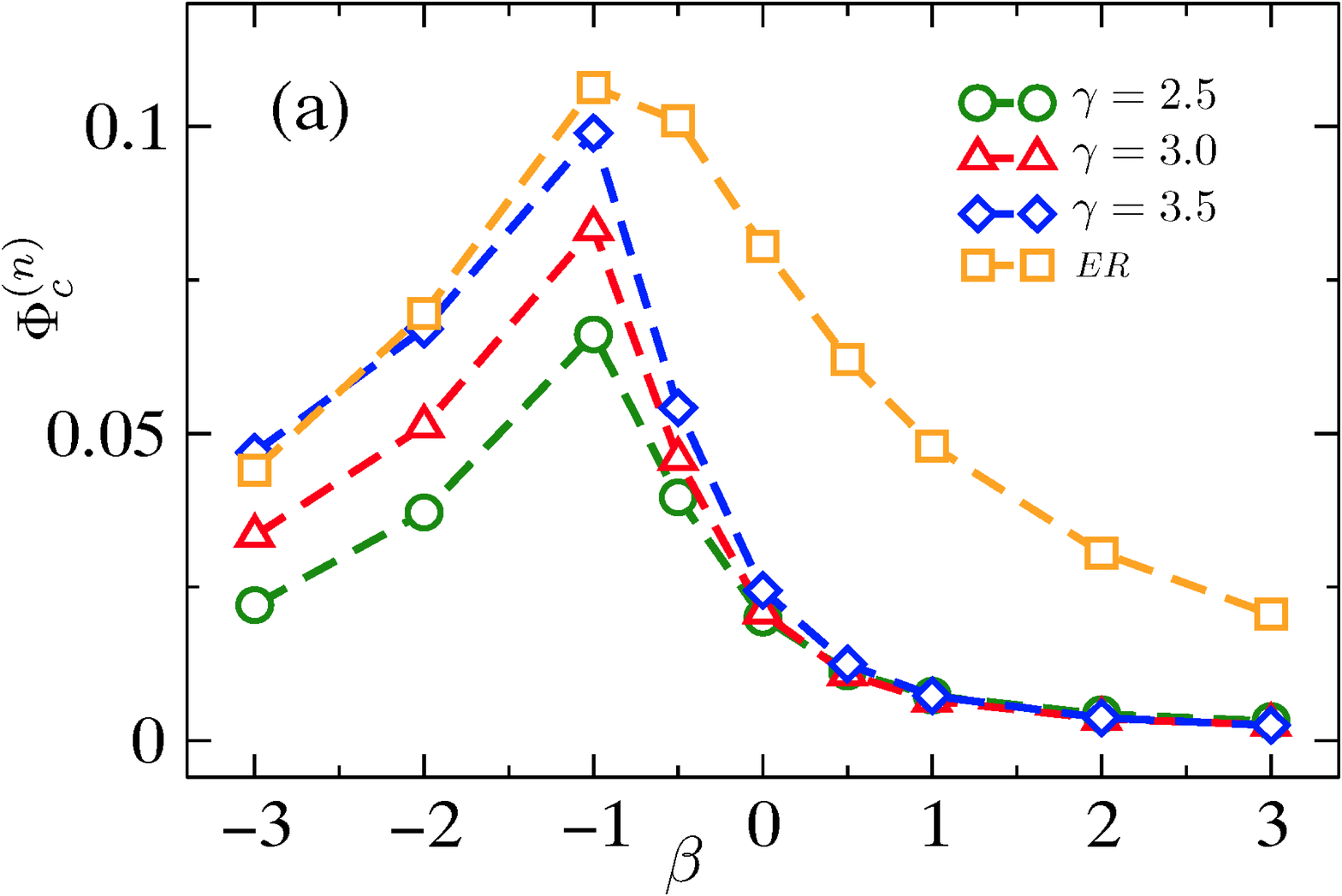}}
\centerline{\includegraphics[width=2.5in]{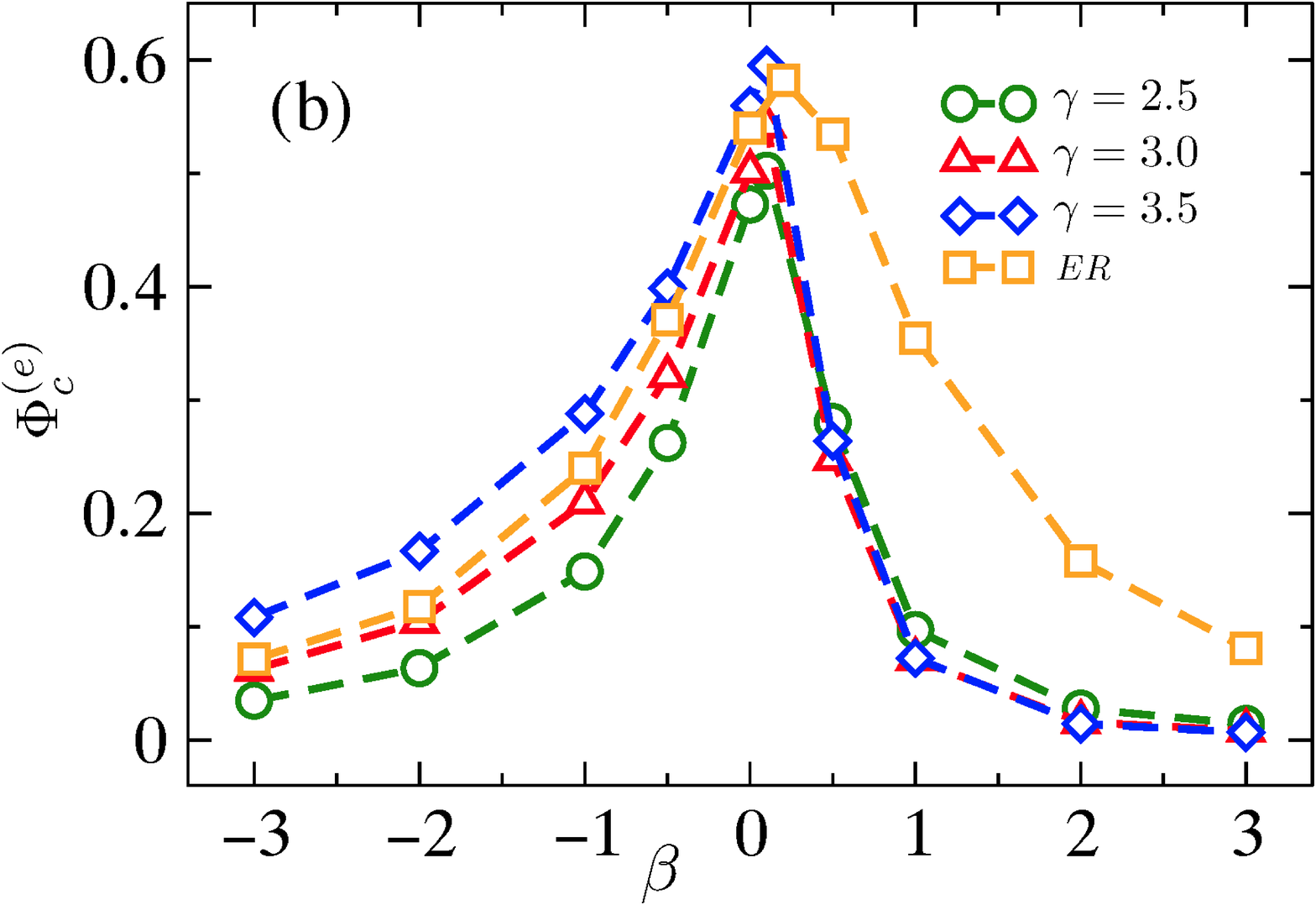}}
\caption{(a) Node-limited throughput and (b) edge-limited throughput is shown as function of $\beta$ control parameter obtained for scale-free and ER networks ($N=10^3$, $\langle k \rangle=10$) averaged over $400$ network realizations. }
\label{fig3}
\end{figure}

Having considered the load landscapes for different $\beta$ ranges, we are interested in how to characterize the flow efficiency in the network. Here, we characterize traffic flow efficiency by network throughput \cite{SameetBottleneck07,KornissBookCh}: the maximum input current $\Phi_c$ that the network is able to transport without becoming congested. To begin with, we consider two cases. In the first case {\em (i)} which we call {\em node-limited}, nodes have identical processing capability, set to unity, and edges have transport capacity (which is also interchangeably referred to as bandwidth) $b$ that is unbounded ($b=\infty$). In the second case {\em (ii)} which we call {\em edge-limited}, the processing capability of every node is unbounded and all edges in the network have a finite and identical bandwidth ($b=1$). In both cases, we now assume that $\Phi$ units of current flow between $N$ source-target pairs simultaneously. As we increase the input current $\Phi$, the node with the highest load will be the first to become congested. Thus, for a given network, we are interested in the problem of choosing the link conductances such that the network is {\em least susceptible to congestion}. To fix notations, we denote by $\Phi_c^{(n)}$ the {\em node-limited throughput}, which is the maximum input current for which the network is congestion free in the node-limited case, and by $\Phi_c^{(e)}$ the {\em edge-limited throughput}, that is the maximum input current for which the network is congestion free in the edge-limited case. As shown earlier \cite{SameetBottleneck07, KornissBookCh}, these quantities are only limited by the maximum vertex load $\ell_{max}$ and maximum edge load $\ell(e)_{max}$ in the network, respectively. Mathematically, they are defined as
\begin{equation}
\Phi_c^{(n)} =\frac{1}{\ell_{max}}, \;\; \Phi_c^{(e)}=\frac{1}{\ell(e)_{max}}\;,
\label{eq:node-edge_limited}
\end{equation}
and their dependence for various link conductances are shown in Figs. \ref{fig3}(a) and \ref{fig3}(b).
For random scale-free networks, for $\gamma$ in the range of $[2.5,
3.5]$, as well as for Erd\H{o}s-R\'{e}nyi (ER) random graphs
\cite{ErdosRgraph}, the node-limited throughput attains a maximum
value for the hub-avoiding flow around $\beta \approx -1$ (observed also previously
in scale-free BA network \cite{Huang_MS2010,KornissBookCh}), contrary to
the edge-limited throughput attaining an optimal value for the
weakly hub-biased flow around $\beta \approx 0.1$ and a slightly higher value around $\beta \approx 0.2$ for ER networks. As it was shown in Figs.
\ref{fig1}(c) and \ref{fig2}(a) the $\beta \approx -1$ and $\beta \approx 0.1$
values are not coincidental. These are the values for which the
vertex and edge loads are balanced, respectively. We also observe
that for all values of $\beta$ and for various network topologies,
$\Phi_c^{(e)}> \Phi_c^{(n)}$, indicating that node-capacity constraints have a
more severe effect on network throughput than bandwidth constraints.
The quantities $\Phi_c^{(n)}$ and $\Phi_c^{(e)}$ increase as the fatness of the
degree distribution tails reduces. Furthermore, the homogeneous
ER graph exhibits a larger node- and edge-limited throughput than
the heterogeneous scale-free topologies with similar
characteristics. Finally, the increase of the network average degree
results in the decrease of the maximum vertex and edge loads, which
in turn contributes to the increase of $\Phi_c^{(n)}$ and
$\Phi_c^{(e)}$ (not shown).

\begin{figure}[htbp]
\centerline{\includegraphics[width=2.5in]{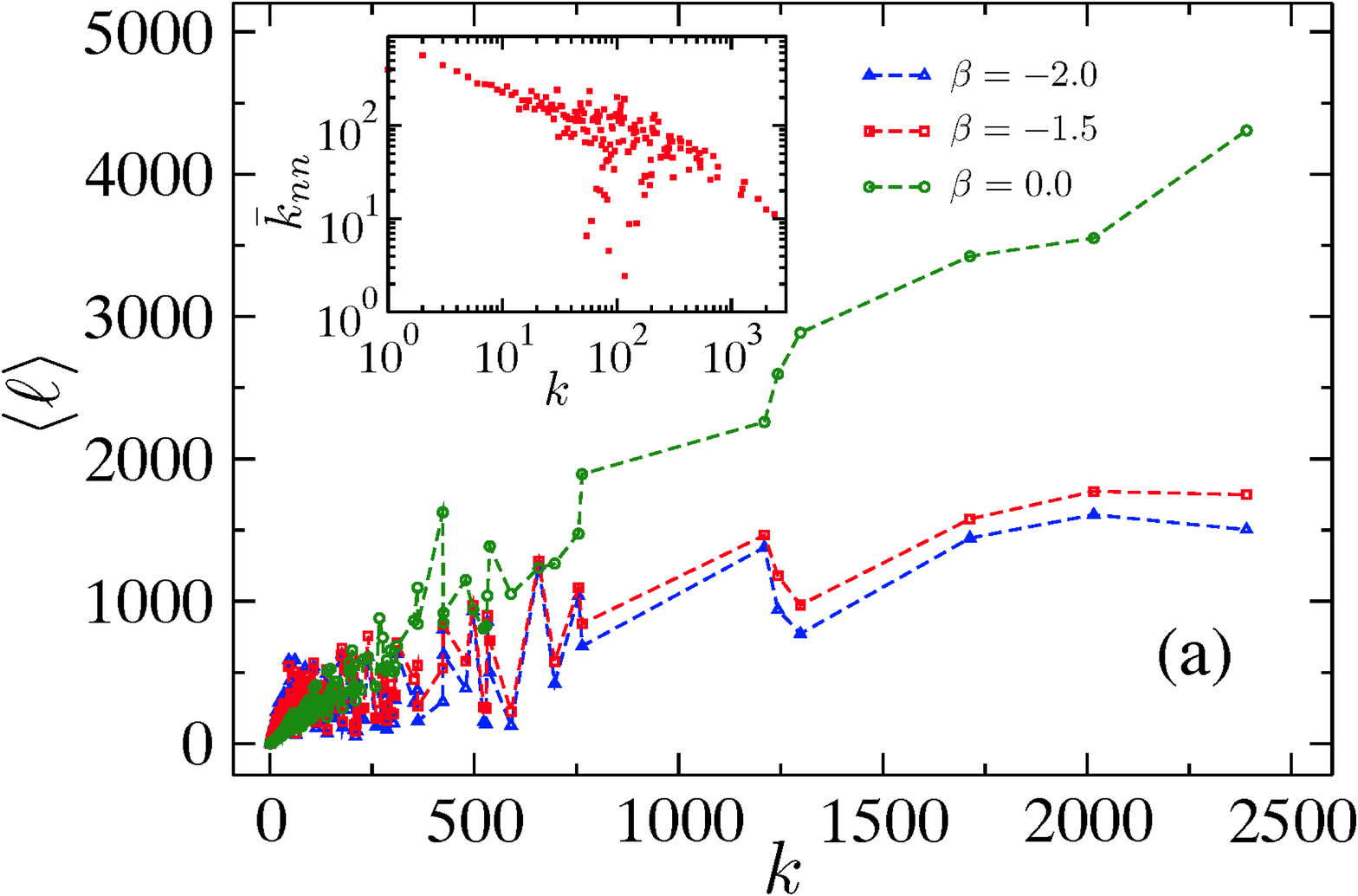}}
\centerline{\includegraphics[width=2.5in]{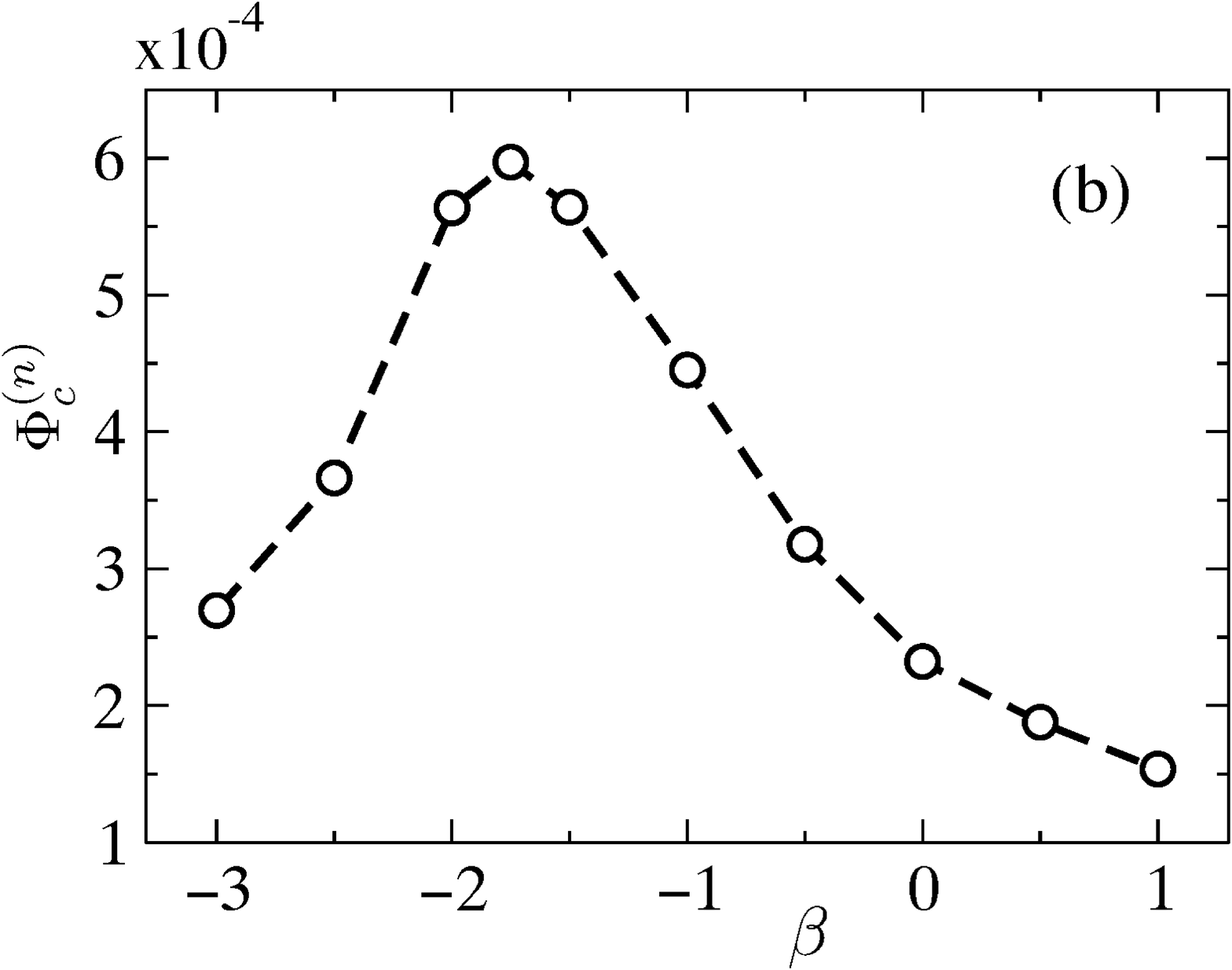}}
\centerline{\includegraphics[width=2.5in]{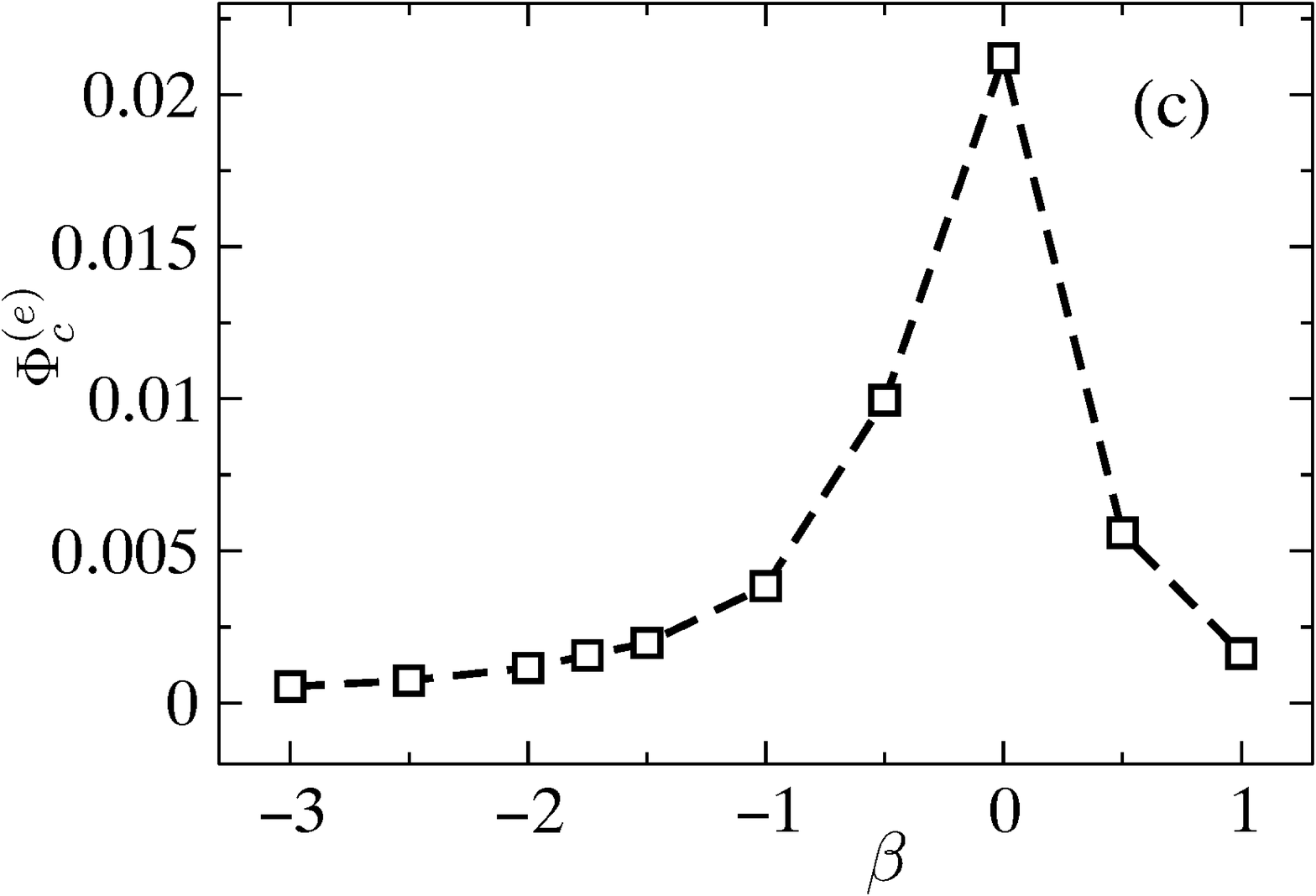}}
\caption{(a) Average vertex load for different degree classes obtained for the Internet at AS level \cite{InternetData06} for hub-avoiding ($\beta=-2.0, -1.5$) and unbiased ($\beta=0.0$) flows. The inset shows the average nearest neighbor degree $\bar{k}_{nn}$ as function of node degree. (b) Node-limited throughput and (c) edge-limited throughput is shown as function of $\beta$ control parameter.}
\label{fig4}
\end{figure}

An interesting application of the above weighted and distributed traffic scheme is its implementation to an actual empirical network structure. To that end, we analyzed a July 2006 snapshot of the Internet at the autonomous system (AS) level \cite{InternetData06}. The network, composed of $22963$ nodes has an average degree $\langle k \rangle =4.22$; it is characterized by a fat-tailed degree distribution and shows disassortative mixing by degree \cite{MixingNewman03}(see inset of Fig. \ref{fig4}(a)).
The vertex load corresponds to the net number of data packets passing through an AS. 
Qualitative similarities are observed between Fig. \ref{fig4}(a) and Fig. \ref{fig1}(b); furthermore between Figs. \ref{fig4}(b), \ref{fig4}(c) and Figs. \ref{fig3}(a) and \ref{fig3}(b), respectively. Following the presented results from our model so far, it is obvious that the vertex loads would be optimally balanced in the hub-avoiding regime. For the Internet, the maximum of the node-limited throughput occurs around $\beta \approx -1.75$ [Fig. \ref{fig4}(b)], a significantly lower value than the one obtained for uncorrelated scale-free networks. Further, the edge-limited throughput is optimal at around $\beta \approx 0$ [Fig. \ref{fig4}(c)], essentially unchanged from uncorrelated scale-free graphs. This behavior of general disassortative networks can be qualitatively understood as follows. Assuming that the scaling of the vertex load is dominated by the weighted degree $C_i=\sum_{j} a_{ij}C_{ij}$ for each node, one can show that the value of $\beta$ which balances the vertex loads will be lower for disassortative networks (such as the Internet) than for uncorrelated ones. Therefore, a stronger hub avoidance is required to balance the loads on a disassortative network as compared to an uncorrelated one. On the other hand, assuming that the edge load is dominated by the weighted couplings $C_{ij}$ [Eq. (\ref{eq:current})], the degree-degree correlations have no major impact scaling of the edge load. Hence, the value of the weighting parameter which balances the edge load remains about $\beta \approx 0$, unchanged from uncorrelated networks. Finally, in both cases one can assume that maximal throughput is associated with balanced loads.

\begin{figure}[htbp]
\centerline{\includegraphics[width=2.5in]{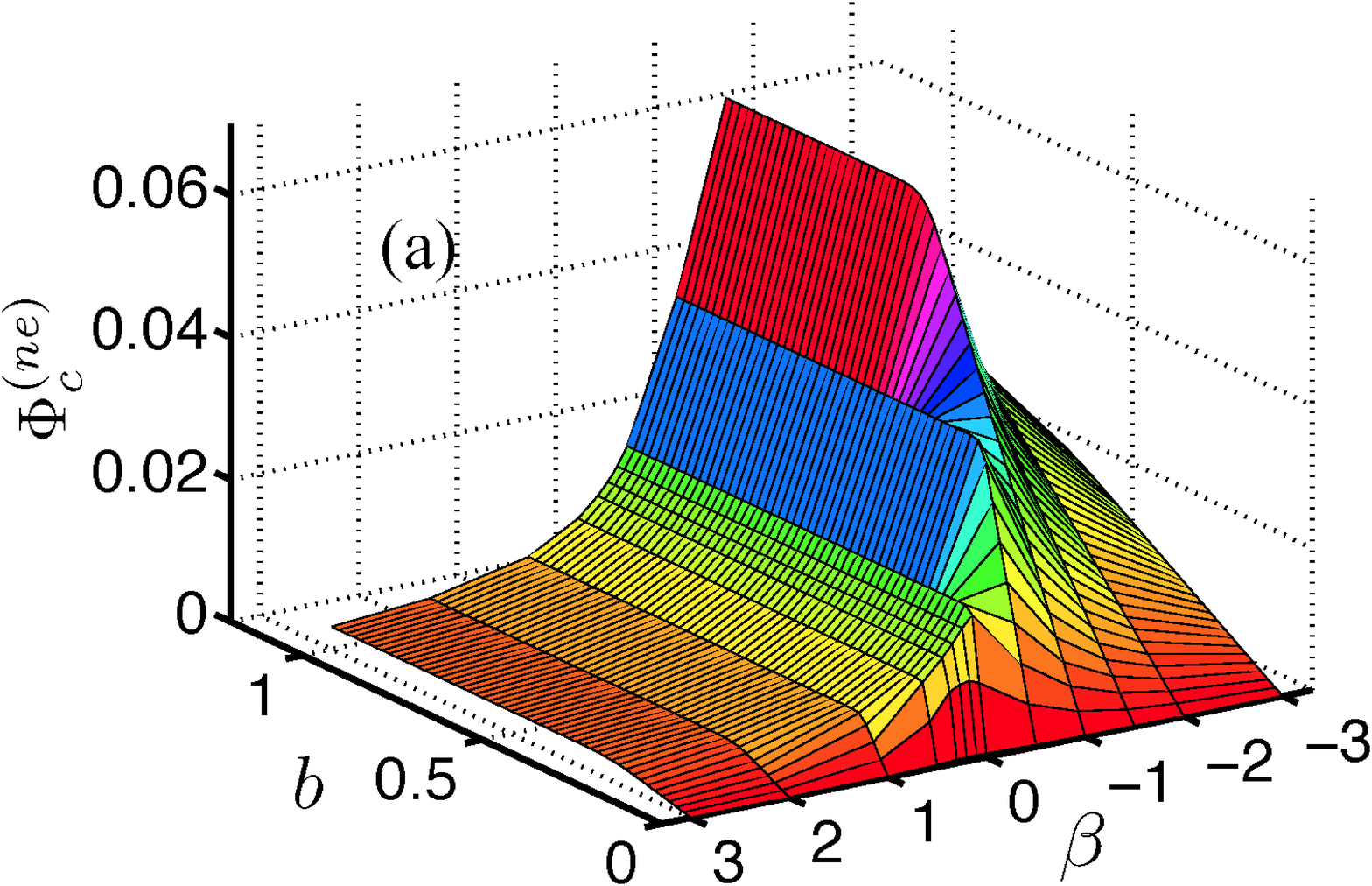}}
\centerline{\includegraphics[width=2.5in]{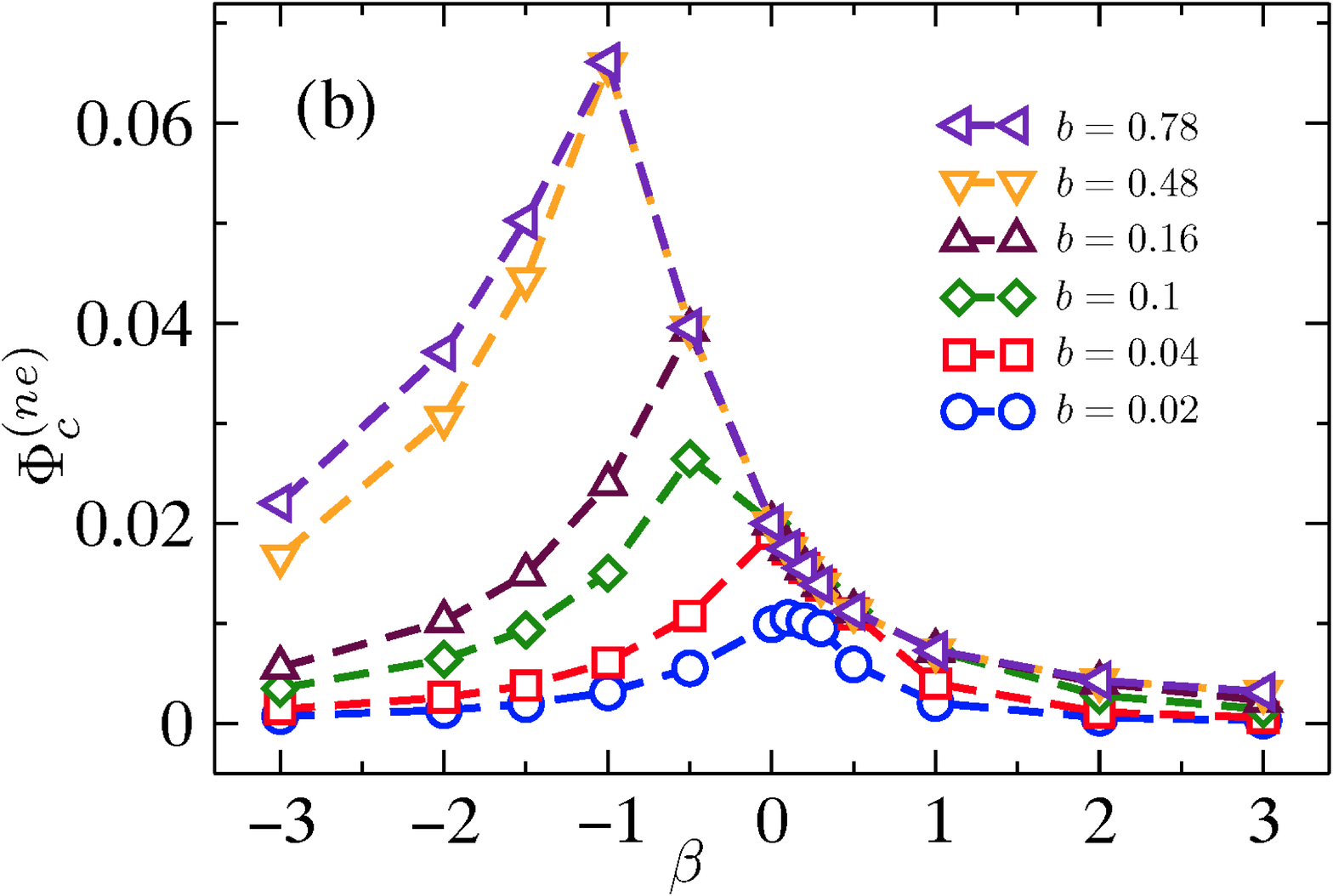}}
\caption{(a) Node-edge-limited throughput $\Phi_c^{(ne)}$ as function of control parameter $\beta$ and finite bandwidth $b$ of edges. (b) Cross-sections of the $\Phi_c^{(ne)}$ surface shown in (a) highlighting the transition of the optimum value from hub-avoiding flows ($\beta \approx  -1$) to weakly hub-biased flows ($\beta \approx 0.1$) as bandwidth is lowered from $b=0.78$ to $b=0.02$. Data were obtained for random scale-free network of degree exponent $\gamma=2.5$. Network specifications are the same as in Fig. \ref{fig1}.}
\label{fig5}
\end{figure}

In addition to the node-limited and edge-limited cases, we consider
a third case {\em (iii)} that we call {\em node-edge-limited}, when all nodes have unit processing
capabilities and all edges have a finite, identical bandwidth, $b
\geq 0$. The network is congestion free for any $\Phi$ units of current flowing
through the network as long as $\Phi \ell_i \leq 1$ {\em and} $\Phi
\ell_{ij} \leq b$ conditions are satisfied for each node $i$ and
edge $(i,j)$, respectively. The {\em node-edge-limited  throughput}
$\Phi_c^{(ne)}$, that is the maximum input current for which the
network is congestion free in the node-edge-limited case is defined
as:
\begin{equation}
\Phi_c^{(ne)} =\frac{1}{\text{max}\{ \ell_i, \ell_{ij}/b\}}\;.
\end{equation}
In the case of infinite bandwidth, $\Phi_c^{(ne)}(b=\infty) \equiv \Phi_c^{(n)}$; for no bandwidth ($b=0$) the network is unable to transfer current flow, therefore $\Phi_c^{(ne)}=0$. As seen in Figs. \ref{fig3}(a) and \ref{fig3}(b) the node capacity constraint $\Phi \ell_i \leq 1$ (for all nodes $i$) is a more restrictive constraint than $\Phi \ell_{ij} \leq 1$ (for all edges $(i,j)$) constraint. Thereby, in this model the unit bandwidth $b=1$ is an upper limit of increasing network throughput expressed mathematically as $\Phi_c^{(ne)} (b=1) \equiv \Phi_c^{(n)}$. By lowering the bandwidth of edges, we restrict the current flowing through the network, consequently decreasing the $\Phi_c^{(ne)}$. This is shown in Fig. \ref{fig5}(a).
As $b$ decreases we observe the continuum of cases between the node-limited and edge-limited extremes. This decrease is accompanied by the shift of the optimal value of $\Phi_c^{(ne)}$ from the region of hub-avoiding flows ($\beta <0$) to the region of slightly hub-biased flows ($\beta >0$) also shown in Fig. \ref{fig5}(b). The overall shape of $\Phi_c^{(ne)}(\beta,b)$ surface remains unaltered as we increase $\gamma$ degree exponent (not shown) in agreement with Figs. \ref{fig3}(a) and \ref{fig3}(b).

\subsection{Heterogeneous allocation of network resources}
\label{Sec4B}
In the previously considered cases we implicitly assumed that the probability of being a source is the same across all nodes $(1)$, and so is the probability of being a target $[1-(1-1/(N-1))^{N-1}]$; additionally, the current flowing between each source-target pair is identical. In the following, we address a particular case of {\em heterogeneous} flow where the incoming and outgoing flow rate  between each source-target pair is proportional to the degree of source $k_s$ and target $k_t$ nodes, namely $\sim (k_s k_t)^{\rho}$, for $\rho \geq 0$ real. Considered previously in \cite{KornissBookCh}, this model of heterogeneous flow is inspired by a study of the world-wide air-transportation network \cite{Barrat04}, wherein the traffic (total number of passengers) handled by each airport was observed to scale as $\sim k^{\theta}$, where $k$ was the total degree of the airport. The appropriately weighted vertex load $\ell_i(\rho)$ in this case is defined as  \cite{KornissBookCh}
\begin{equation}
\ell_i(\rho) =\frac{N}{\sum_{s,t} (k_s k_t)^{\rho}} \sum_{s,t} (k_s k_t)^{\rho} I_i^{st}, \quad \forall i=1,\dots,N.
\end{equation}
Note, $\ell_i(\rho=0)=\ell_i$. Further, we denote by $\Phi_c^{(n)}(\rho)$, the {\em node-limited weighted throughput} and by $\Phi_c^{(e)}(\rho)$ the {\em edge-limited weighted throughput} that are generalizations of the node-limited and edge-limited throughputs defined in the previous subsection. Therefore, similar to $\Phi_c^{(n)}$ and $\Phi_c^{(e)}$ defined by (\ref{eq:node-edge_limited}), their weighted counterparts are only limited by the maximum weighted vertex and edge loads, respectively.

\begin{figure}[htbp]
\centerline{\includegraphics[width=2.5in]{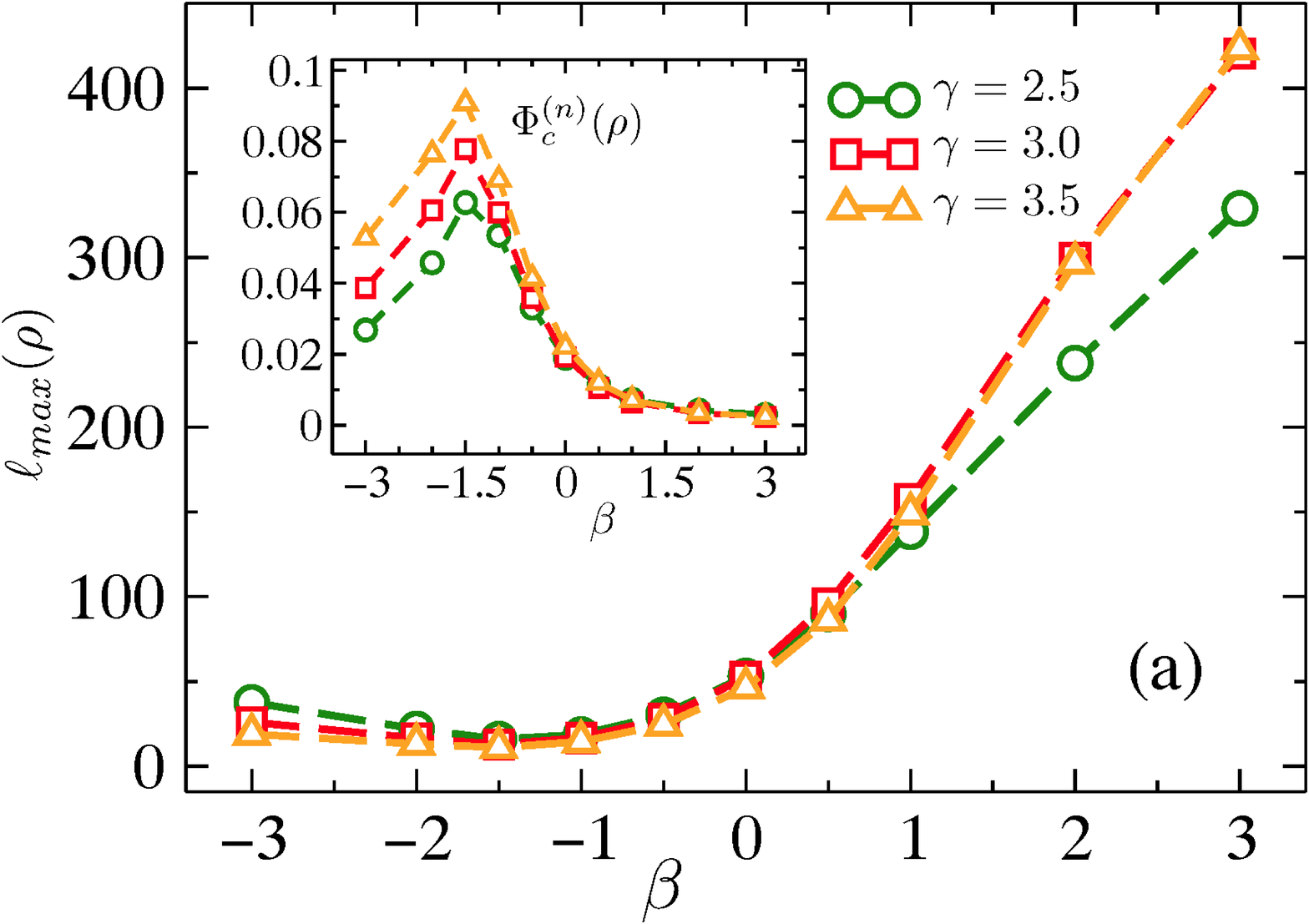}}
\centerline{\includegraphics[width=2.5in]{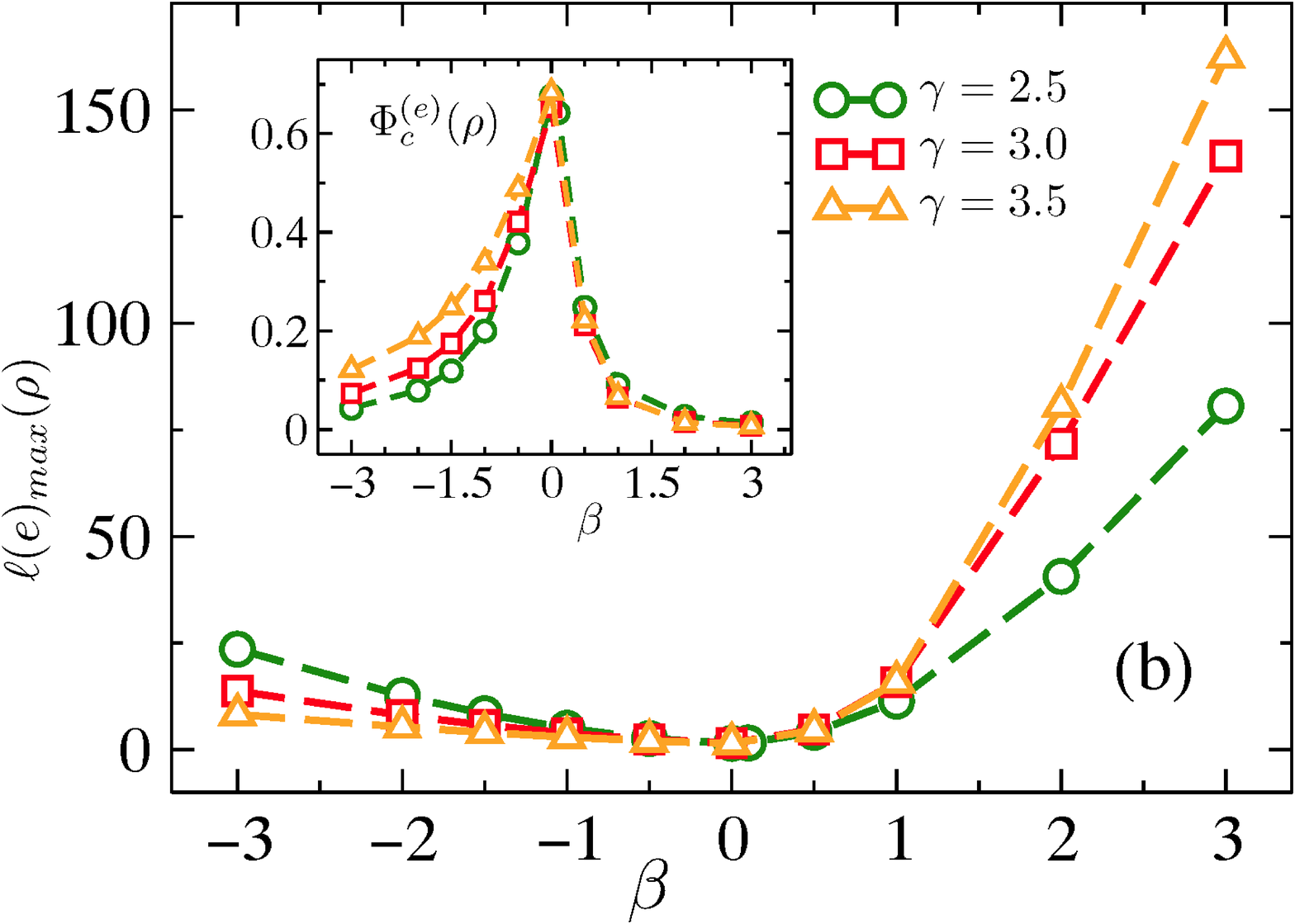}}
\caption{(a) Maximum weighted vertex load and (b) maximum weighted edge load as function of $\beta$ control parameter for $\rho=1.0$. Data were obtained for scale-free networks ($N=10^3$, $\langle k \rangle=10$) with various $\gamma$ degree exponents, averaged over $400$ network realizations. Insets show the node-limited weighted throughput $\Phi_c^{(n)}(\rho)$ and edge-limited weighted throughput $\Phi_c^{(e)}(\rho)$, respectively, when $\rho=1.0$.}
\label{fig6}
\end{figure}

In the following, we restrict our studies to the case when $\rho=1.0$. Qualitative similarities are observed between the insets of Figs. \ref{fig6}(a), (b) and Figs. \ref{fig3}(a) and \ref{fig3}(b), respectively.
Additionally, we observe a shift towards lower values of $\beta$ in the position of both node-limited and edge-limited optimal throughputs. Specifically, the $\Phi_c^{(n)}(1.0)$ attains a maximum value for the hub-avoiding flow around $\beta \approx -1.5$, while $\Phi_c^{(e)}(1.0)$ reaches its maximum value for the unbiased flow, when $\beta \approx 0$. The strong hub avoidance arises to compensate for the hub bias implicitly introduced by the heterogeneous flow ($\sim k^{\rho}$).
 The network throughput for this case is significantly lower than for the case of $\rho=0$.

\begin{figure}[htbp]
\centerline{\includegraphics[width=2.5in]{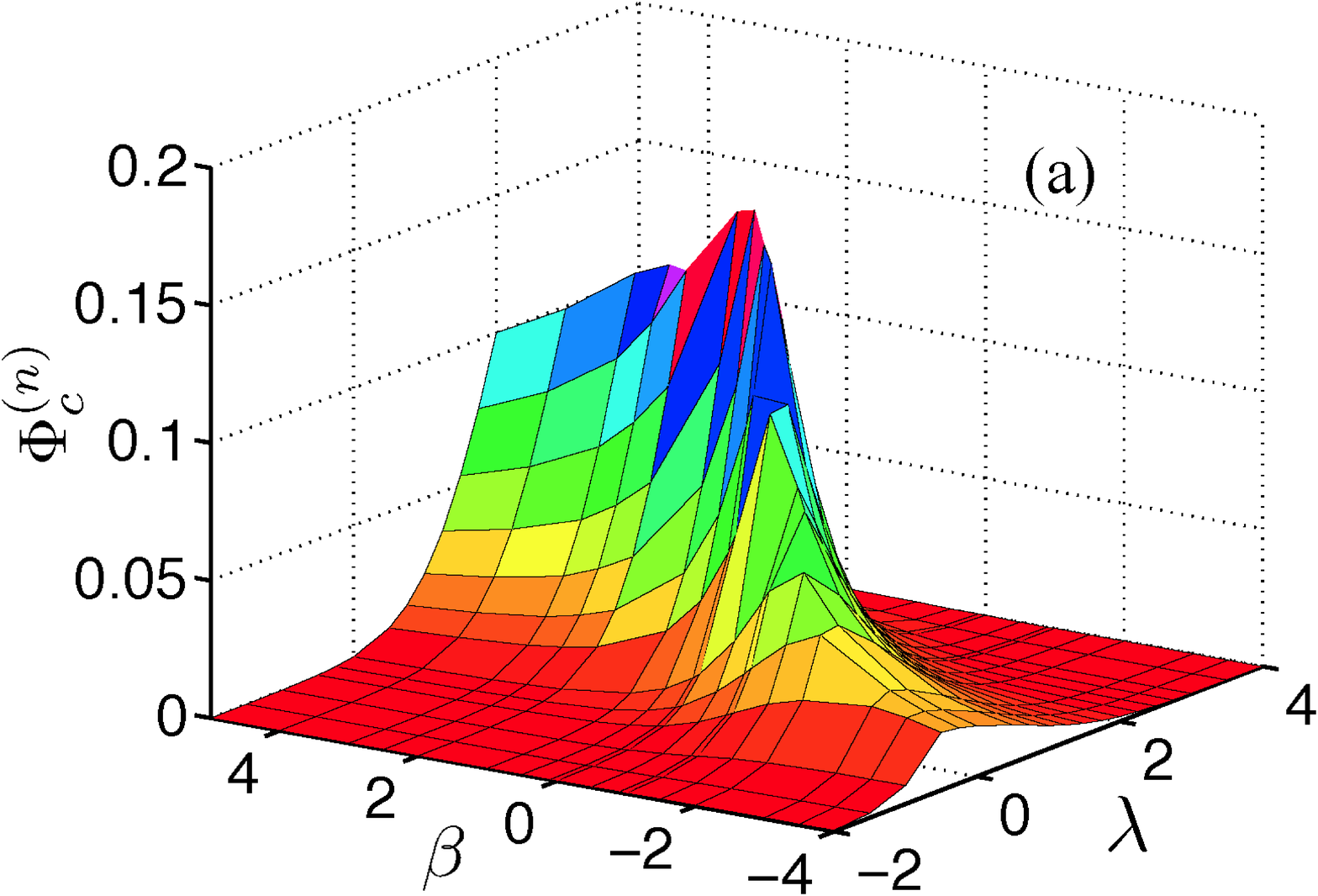}}
\centerline{\includegraphics[width=2.5in]{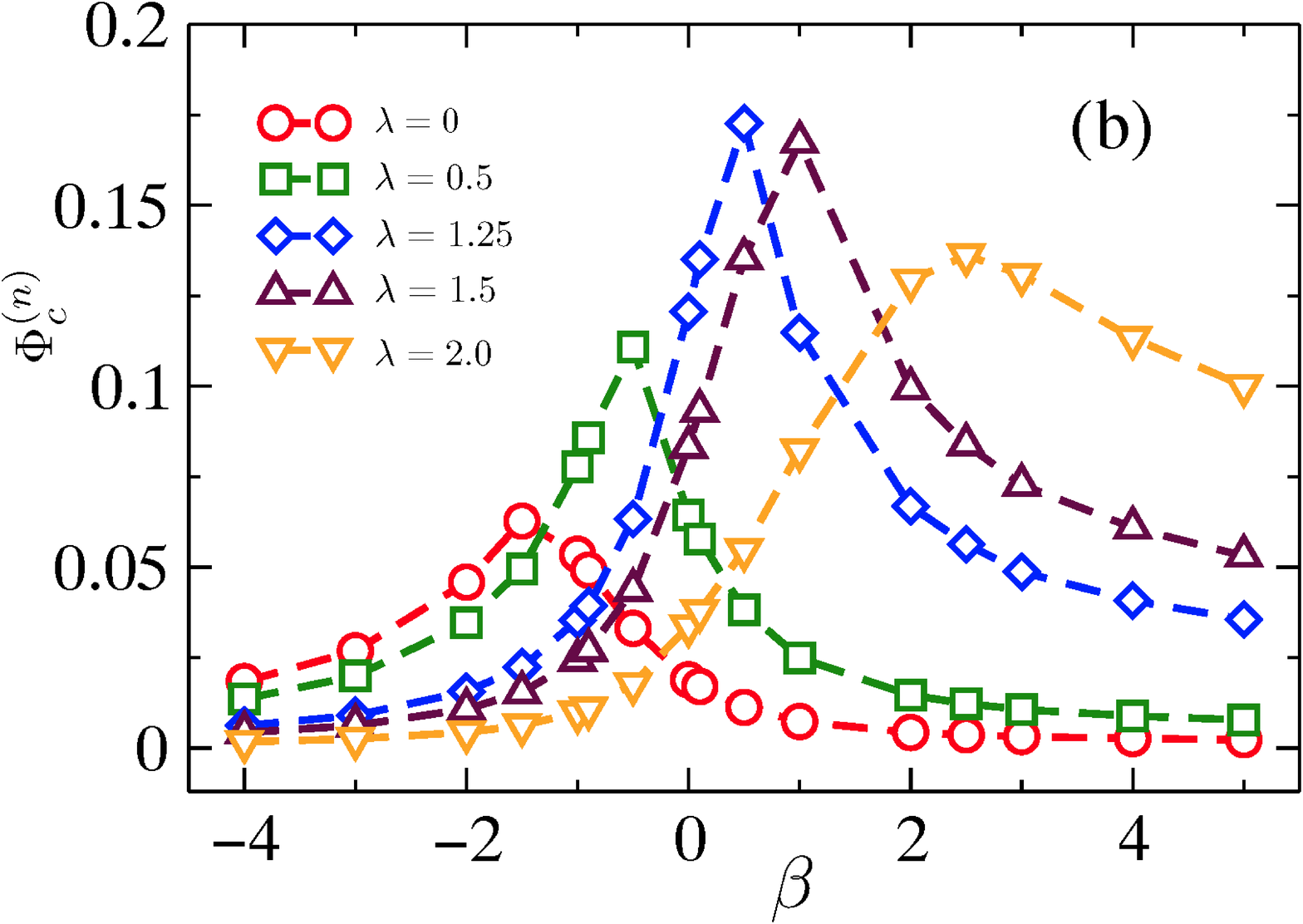}}
\caption{(a) Node-limited weighted throughput when nodes' processing capabilities follow the parametrized form of $Q_i(\lambda) \sim k_i^{\lambda}$ where $\lambda$ is a real parameter and $\rho=1.0$. (b) Cross-sections of throughput surface shown in (a) highlighting the relevant range of $\lambda$ values for which the throughput is optimal. Data were obtained for scale-free networks of $\gamma=2.5$. Network specifications are the same as in Fig. \ref{fig1}.}
\label{fig7}
\end{figure}

Further on, we restrict ourselves to the case when there is infinite bandwidth associated to the edges and the only flow limiting factor is the processing capacity of the nodes, denoted by $Q_i$ for all node $i$ in the network. Rather than consider an optimization of the functional form of the $Q$ distribution, we consider the case when $Q_i$ follows  a parametrized form $Q_i (\lambda) \sim k_i^{\lambda}$. As previously, we consider the constraint that the sum of the node processing capacities is equal to $N$. We then focus on the question of how the node processing capacities can be distributed such that the congestion-free throughput of the network is maximized. The network is congestion free as long as condition
\begin{equation}
\Phi \ell_i(\rho) \leq \frac{k_i^{\lambda}N}{\sum_{j}k_j^{\lambda}}, \;\; \forall i=1,\dots, N
\end{equation}
is satisfied. Fig. \ref{fig7}(a) depicts the $\Phi_c^{(n)}(\rho=1.0)(\beta, \lambda)$ surface. A clearer picture of the relevant range of $\lambda$ values for which this particular network throughput becomes optimal is shown in Fig. \ref{fig6}(b).
The global maximum of the throughput is attained for hub-biased flow $(\beta \approx 0.5)$ when  $\lambda \approx 1.25$. As $\lambda$ is increased, not surprisingly, optimal throughputs occur at increasing values of $\beta$, i.e. for increasingly hub-biased flows. However optimal throughput values themselves behave non-monotonically, increasing as $\lambda$ is increased from zero and then decreasing after a maximum is attained around $\lambda \approx 1.25.$ The scale-free world wide airline transportation network \cite{Barrat04} has been reported to operate in the hub-biased regime. However, to make any conclusive statements whether this network operates close to its optimal regime would be debatable for lack of empirical data on how the flow of passengers between source-target airports as well as airport capacities scale with local connectivity.

\subsection{Comparing distributed flow to shortest path flow}
\label{Sec4C}
\begin{figure}[htbp]
\centerline{\includegraphics[width=2.5in]{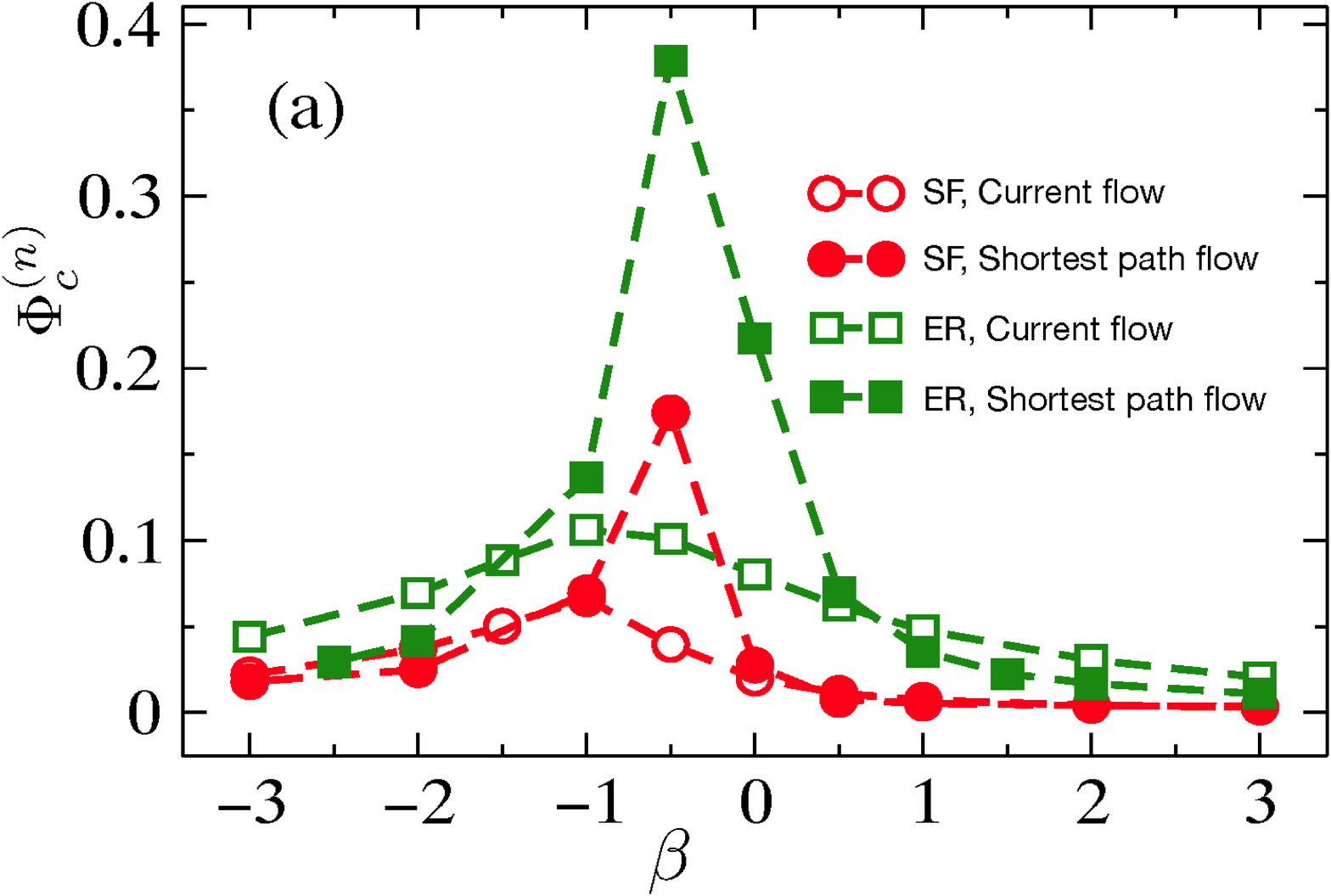}}
\centerline{\includegraphics[width=2.5in]{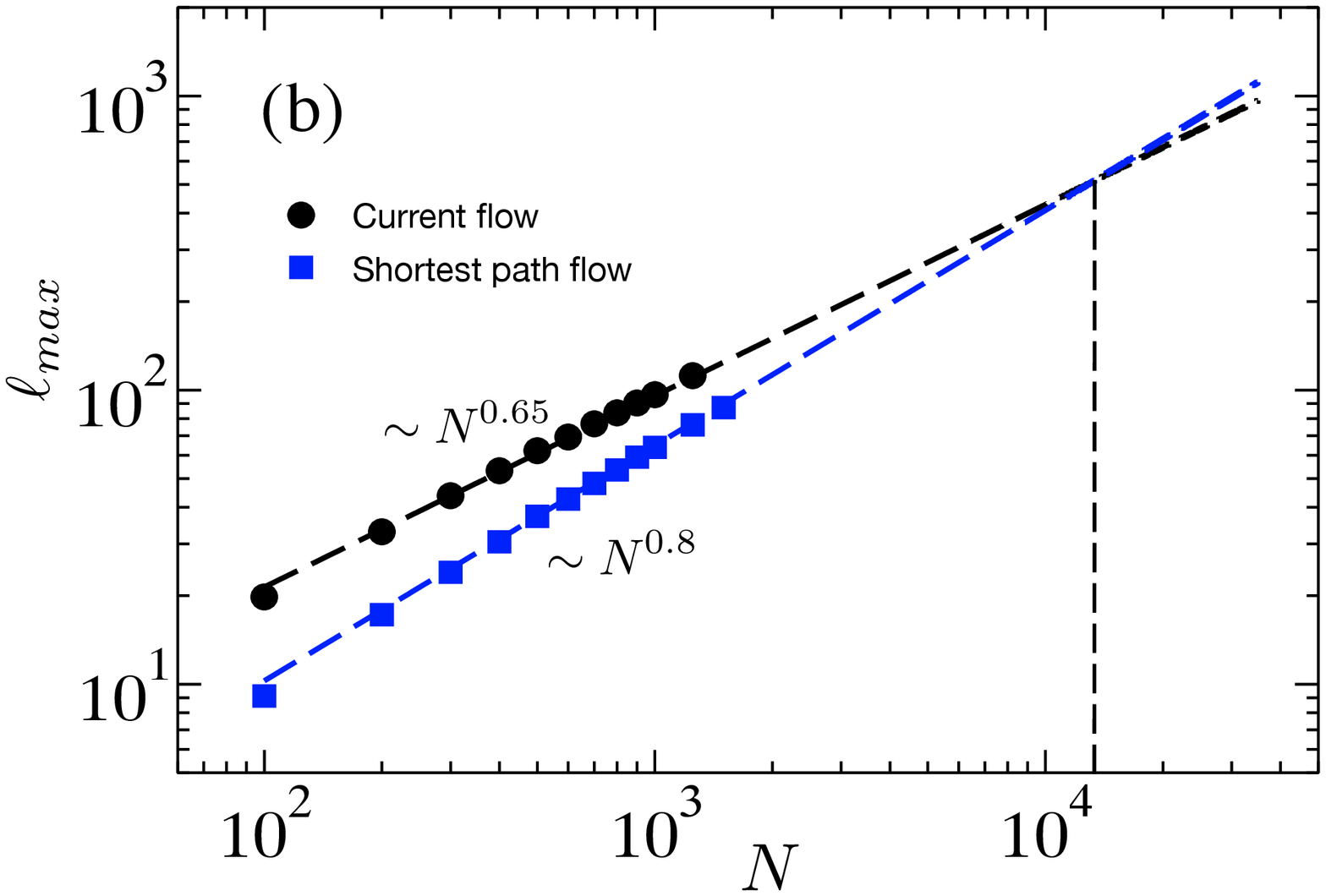}}
\caption{(a) Node-limited network throughput obtained for two different flow strategies: current and shortest-path flow, for scale-free ($\gamma=2.5$) and ER networks ($N=10^3$, $\langle k \rangle=10$). (b) System size dependence of the largest vertex load on unbiased ($\beta=0.0$) current (black circles) and shortest path flow (blue squares) in scale-free networks ($\langle k \rangle =5.0$).}
\label{fig8}
\end{figure}

Our study has focused on distributed flows that utilize all possible paths between a source and target node pair  in proportion to the conductances of these paths. As pointed out in the introduction, another commonly studied model of flow is one that utilizes solely the shortest path(s) between a source/target node pair. In the latter case the relevant measure of load for the optimization problem we discuss here is the shortest-path betweenness centrality \cite{Freeman77,Brandes08}. We now compare the two different flow strategies to examine which one would yield a higher network throughput. The node-limited throughput $\Phi_c^{(n)}$, defined in Sec. \ref{Sec4A} is plotted for random scale-free and ER networks in Figs. \ref{fig8}(a).
Here, the weight associated to edge $(i,j)$ is $1/C_{ij}$ and the shortest-path betweenness centrality is defined with a similar normalization as in the case of current-flow betweenness: $C_B(i)=1/(N-1) \sum_{s \neq t \neq i} \sigma_{st}(i)/ \sigma_{st}$ for any node $i$. $\sigma_{st}(i)$ denotes the number of those shortest paths between source $s$ and target $t$ node that pass through node $i$, while $\sigma_{st}$ gives the total number of shortest paths between $s$ and $t$. The maximum of $\Phi_c^{(n)}$ for both types of flows is attained in the hub-avoiding regime, although the current flow requires a stronger hub avoidance ($\beta \approx -1$) than shortest-path flow ($\beta \approx -0.5$). The optimal values for both cases are larger in case of ER graphs: the homogeneous structure of the network contributes positively to its transport ability. As seen from Fig. \ref{fig8}(a), for the particular system size $N=10^3$, the maximum value of node-limited throughput is larger in the case of shortest paths than for currents ($\Phi_c^{(n)}(sh.p.) > \Phi_c^{(n)}(current)$). To see how this comparative feature is affected by network size, we investigated the values of $\ell_{max}$ on networks of different sizes, for both kinds of flow. The results shown in Fig. \ref{fig8}(b) indicate that shortest path betweenness scales faster with $N$ than current flow betweenness. Extrapolation of the linear fits indicate that this trend ($\Phi_c^{(n)}(sh.p.) > \Phi_c^{(n)}(current)$) holds up to $15000$ nodes, after which $\Phi_c^{(n)}(sh.p.)<\Phi_c^{(n)}(current)$. The maximum vertex load scales with system size as $\sim N^{0.65}$ while the highest shortest path betweenness centrality scales with system size as $\sim N^{0.80}$, value also reported in \cite{SameetBottleneck07}.

\section{Robustness against cascading failures}
\label{Sec5}
Cascading failures embody one of the common vulnerabilities of infrastructure networks \cite{Congestion88, Sachs09}. Models of cascading failures have been previously studied in \cite{MorenoEPL02, AdilsonCasc02, HolmeattackPre02, CrucittiPre04}. Here we study the model of cascading failures introduced by Motter and Lai \cite{AdilsonCasc02}, extending it to the case of distributed flows. As defined in \cite{AdilsonCasc02}, the initially assigned processing capacities of nodes are $Q_i=(1+\alpha)\ell_i^{(0)}$, where $\ell_i^{(0)}$ is the current vertex load on node $i$ for unit current flowing between all source-target pairs as defined in Sec. \ref{Sec2}, for an undamaged network. Equivalently, each node $i$ can handle an excess load $\alpha \ell_i^{(0)}$ where $\alpha \geq 0$ is called tolerance (excess capacity) parameter. A cascading failure is then triggered by the removal of node $i^*$ with highest load. The loss of node $i^*$ results in the redistribution of flows through the network that further leads to the redistribution of loads in the network. By assuming that failure happens at all nodes whose loads are greater than their respective capacities, the (simultaneous) removal of all failed nodes results in the redistribution of loads among the remaining nodes, thereby resulting in more overloaded nodes and more failures. The cascade ends when none of the vertex loads exceed their respective capacities.

In the following we analyze cascades triggered by the removal of the highest load and ask for what values of conductance $\beta$ weighting parameter and $\alpha$ excess capacity parameter is the network the most resilient to such cascades. Following \cite{AdilsonCasc02}, the resilience of a network is quantified in terms of the fraction of surviving largest connected component after the cascade: $G=N'/N$, where $N'$ is the number of nodes belonging to the largest network component after the cascade and $N$ is the undamaged (connected) network size.

\begin{figure}[!htbp]
\centerline{\includegraphics[width=2.5in]{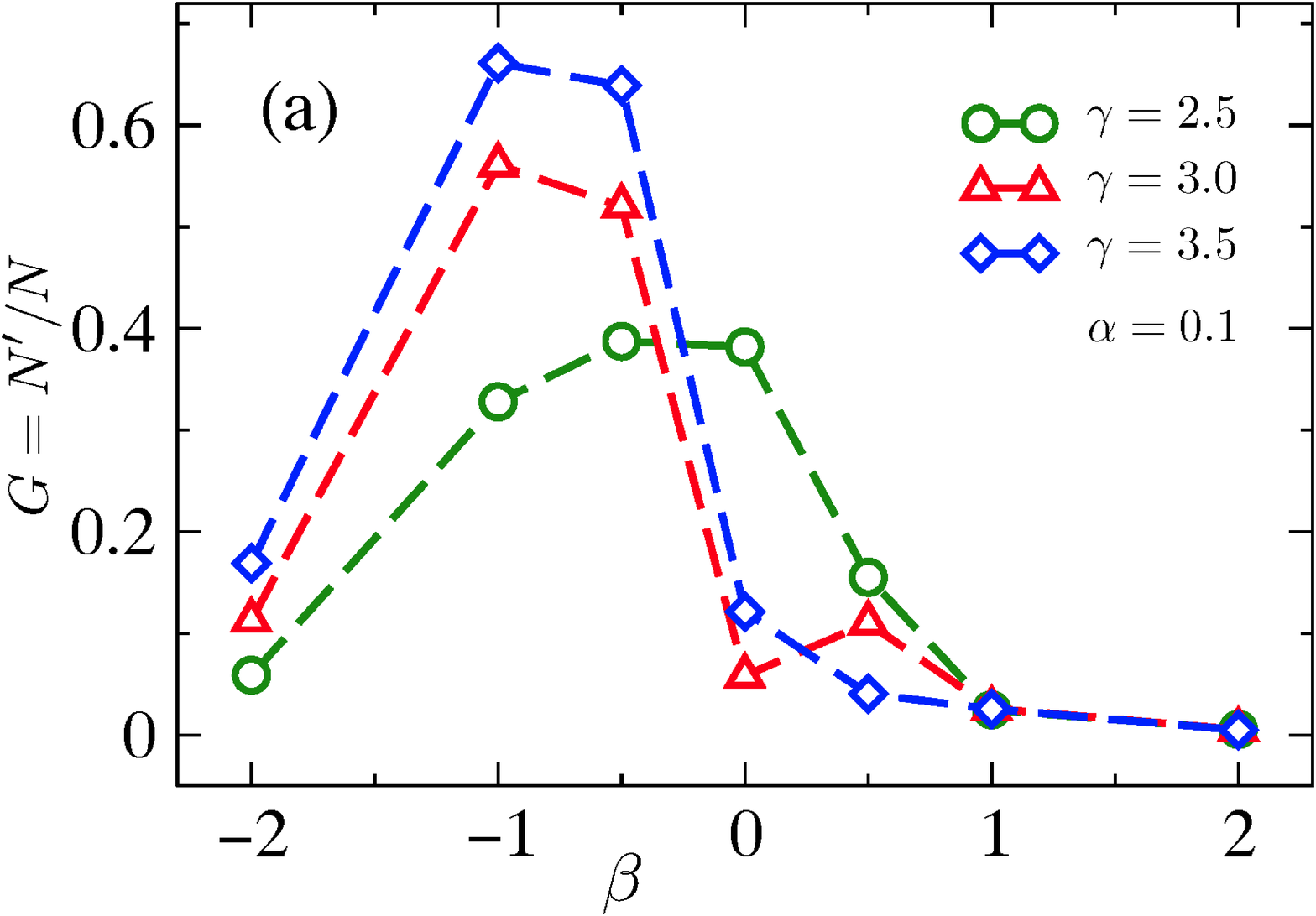}}
\centerline{\includegraphics[width=2.5in]{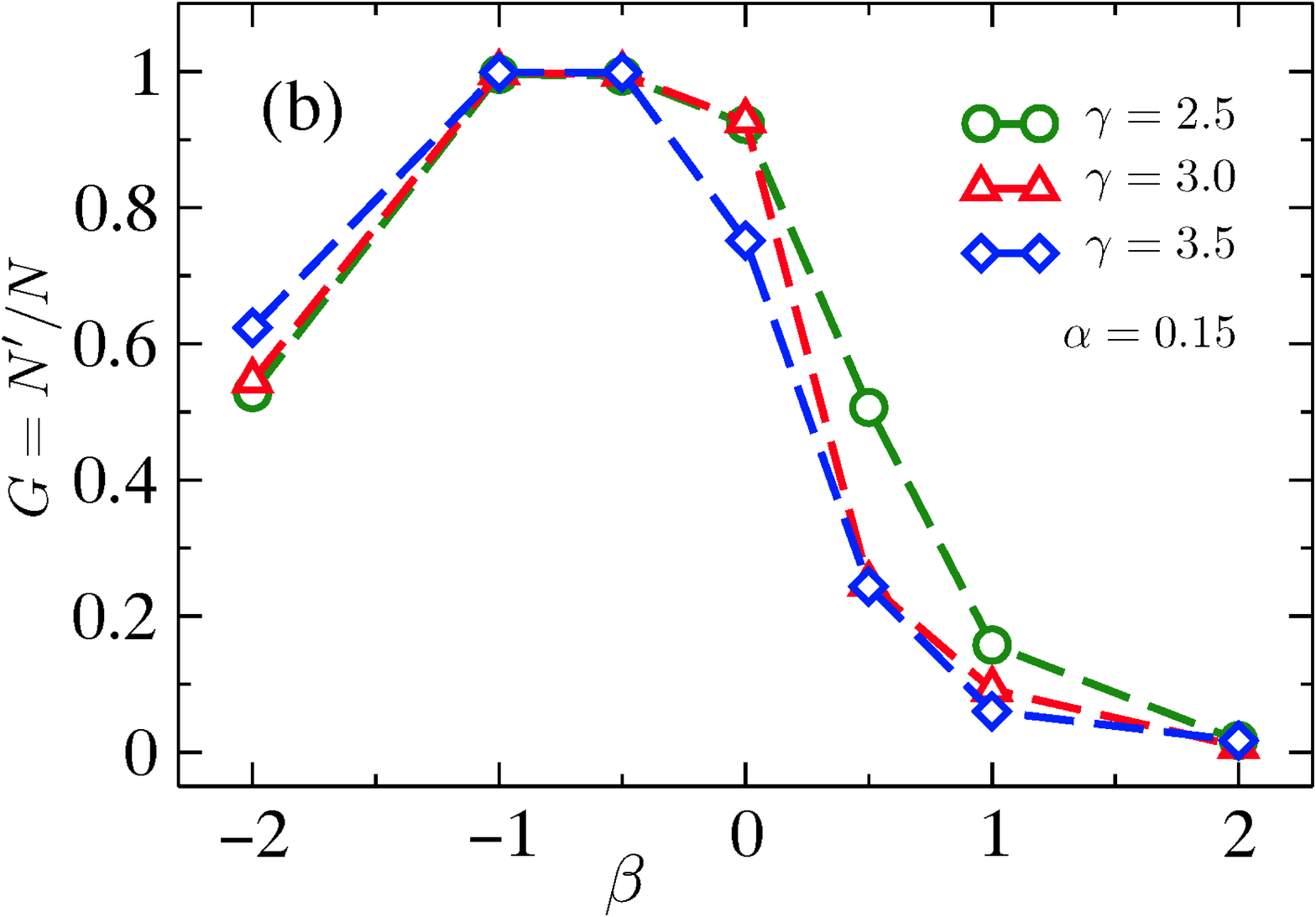}}
\caption{Fraction of the largest surviving network component following a cascading failure ($G$) triggered by the removal of the highest vertex load as function of $\beta$ conductance weighting parameter for (a) $\alpha=0.10$ and (b) $\alpha=0.15$ tolerance parameters. Data were obtained for scale-free networks ($N=10^3$, $\langle k \rangle=4.5$) averaged over $60$ network realizations.}
\label{fig9}
\end{figure}

Network vulnerability to cascades can be increased by increasing the excess load capacity of nodes. However, this might not always be a feasible, cost-effective solution. The question, therefore arises: is there an optimal conductance weighting scheme of the network that would limit the damage resulting from a cascading failure for low values of $\alpha$? As shown in Figs. \ref{fig9}(a) and \ref{fig9}(b) the network resilience $G$ against cascading failure for low values of $\alpha$ attains its maximum for hub-avoiding flows around $\beta \approx -1$, the same value for which the node-limited network throughput, shown in Sec. \ref{Sec4A}, becomes the highest.
Hence, a {\em necessary} condition for scale-free networks (including the BA model) to achieve the highest resilience against cascading failures is having a balanced vertex load profile.  This condition also assures that the node-limited network throughput is optimal (see Sec. \ref{Sec4A}). In general, hub-avoiding flows result in relatively higher network resilience to cascades than hub-biased flows.

\begin{figure}[!htbp]
\centerline{\includegraphics[width=2.5in]{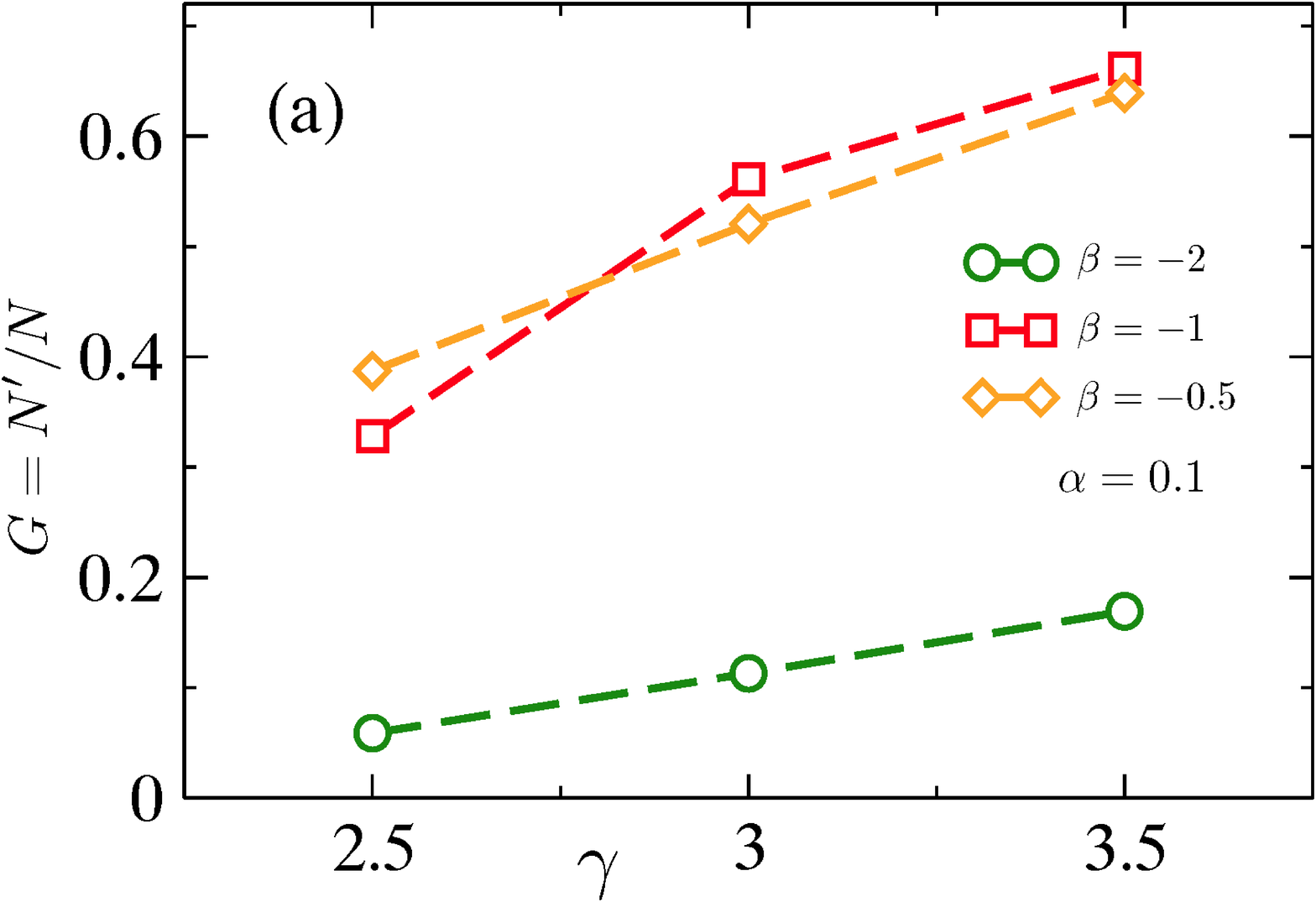}}
\centerline{\includegraphics[width=2.5in]{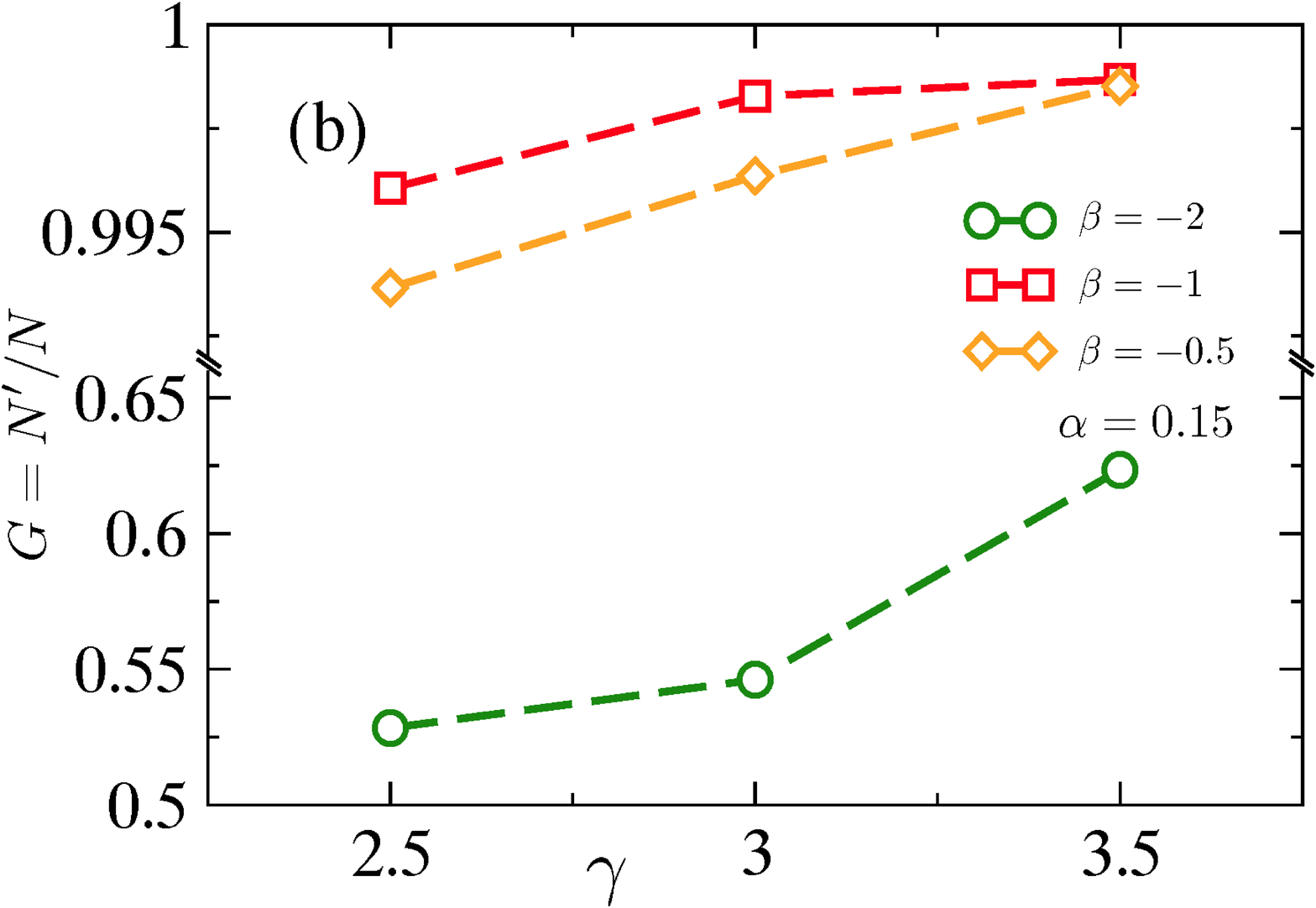}}
\centerline{\includegraphics[width=2.5in]{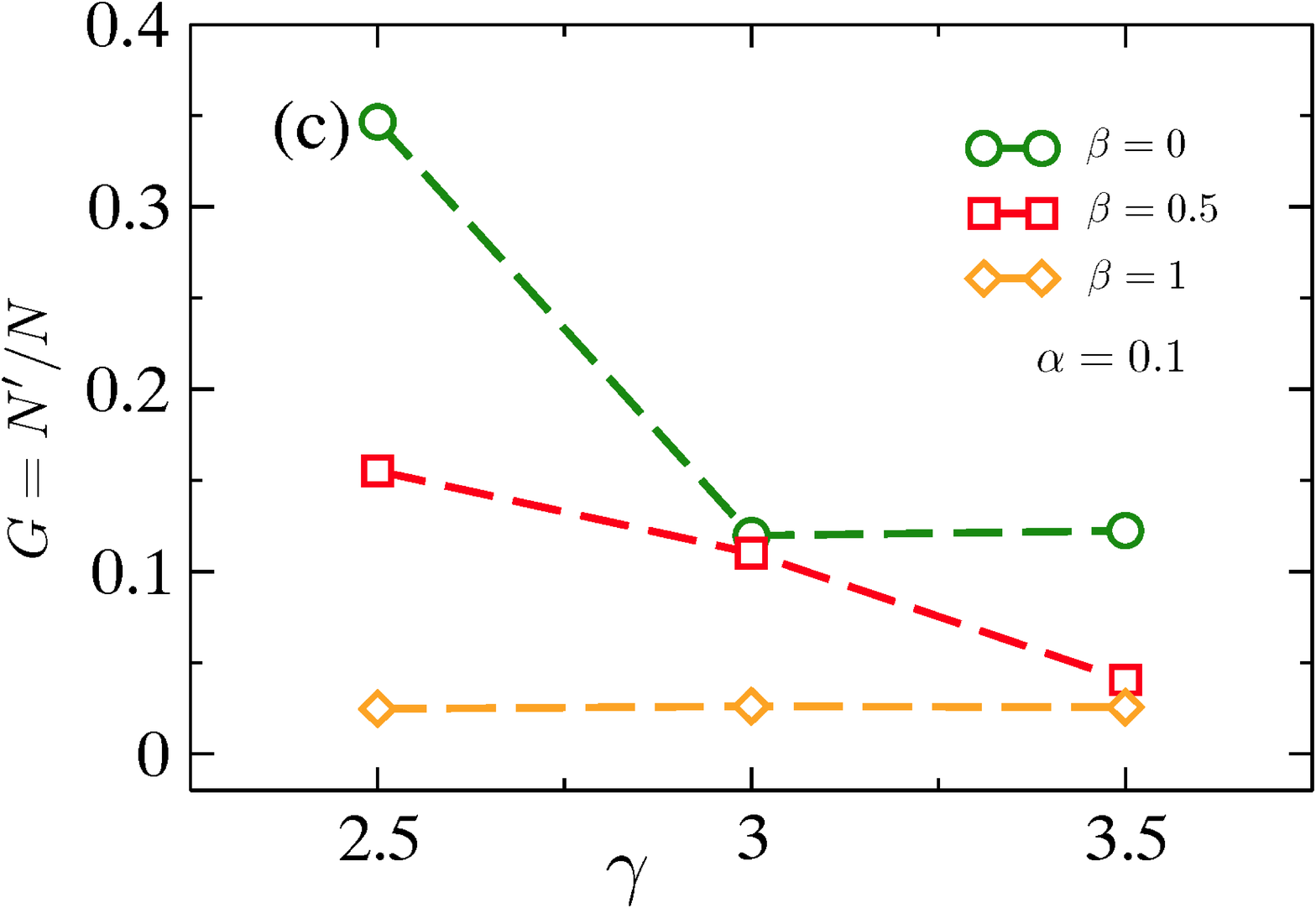}}
\centerline{\includegraphics[width=2.5in]{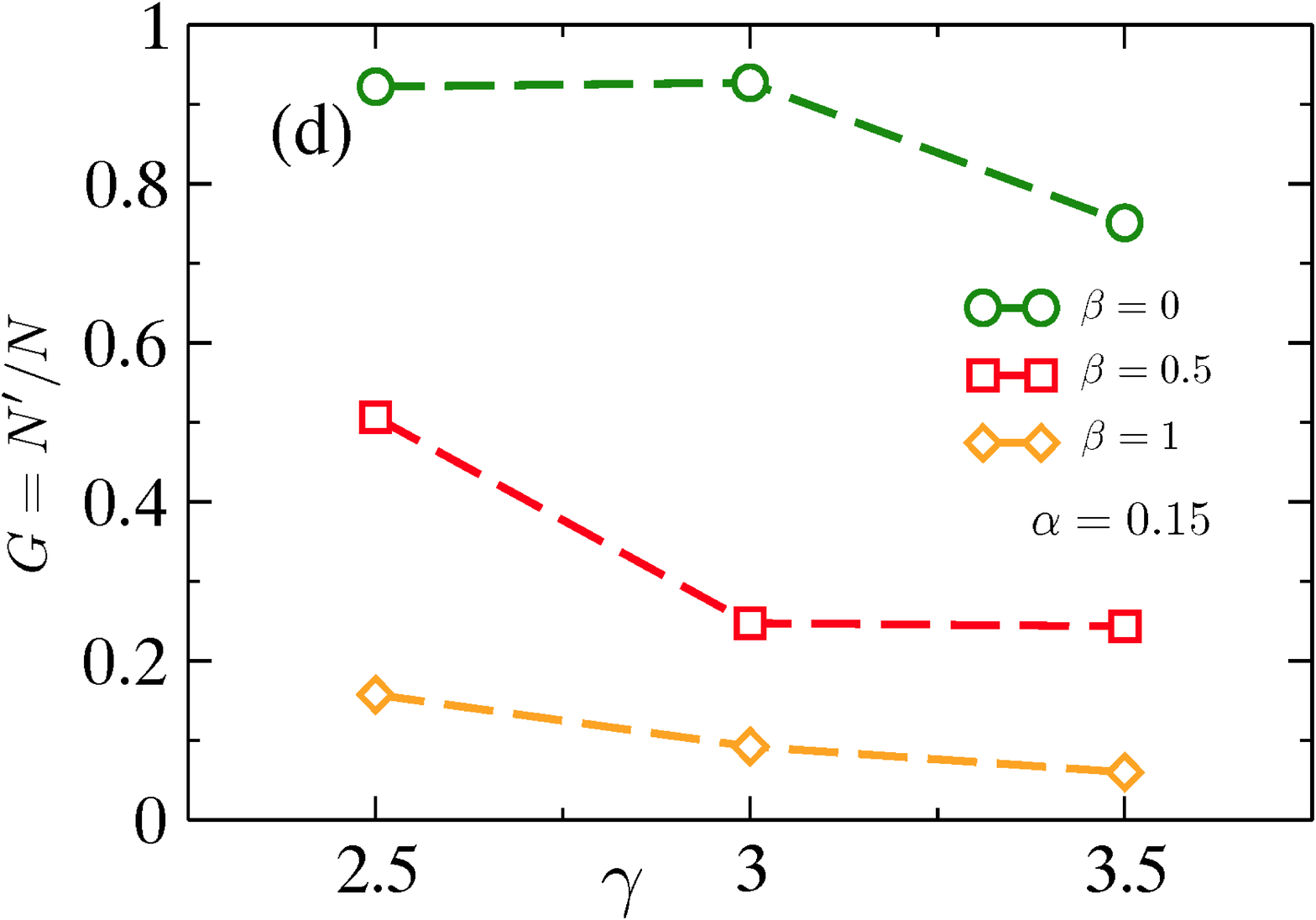}}
\caption{Fraction of the largest surviving network component after a cascading failure ($G$) in scale-free networks ($N=10^3$, $\langle k \rangle =4.5$) as function of $\gamma$ degree exponent. (a) and (b) correspond to the case of hub avoiding flows while (c) and (d) correspond to the case of unbiased and hub-biased flows. Data is shown for two values of the tolerance parameter: $\alpha=0.1$ ((a) and (c)) and $\alpha=0.15$ ((b) and (d)). Error bars are smaller than symbol size.}
\label{fig10}
\end{figure}

As Fig. \ref{fig9} already points out, the network resilience against cascading failures varies with the degree exponent of the network. This is shown in detail in Fig. \ref{fig10} for two, relatively low values of $\alpha$ ($0.1$ and $0.15$.)
For hub avoiding flows [Figs. \ref{fig10}(a), (b)], increasing $\gamma$ yields more robust structures; for unbiased and hub biased flows [Figs. \ref{fig10}(c), (d)], increasing $\gamma$ yields less robust structures. In the latter case, the cascade is triggered by the removal of a high degree node [Fig. \ref{fig1}(b)] which in turn results in the overload of many low degree nodes. The failure of low degree nodes results in a larger damage of a scale-free network with narrow degree distribution, i.e., $\gamma=3.5$ compared to a network with broader degree distribution, i.e., $\gamma=2.5$. In the case of hub-avoiding flows, the cascade is triggered by the removal of a low degree node [Fig. \ref{fig1}(b)] resulting in the overload of some of the large degree nodes. The removal of hub(s) yields a larger connected component in a scale-free network with narrow $P(k)$ compared to a network with broader $P(k)$. The monotonicity of function $G(\gamma)$ for all values of $\beta$  indicates that network robustness is solely determined by network topology and link conductances for a fixed value of $\alpha \ell_i^{(0)}$ excess load capacity.

\section{Summary}
\label{Sec6}
We have studied the problem of optimizing the throughput and robustness of networks carrying distributed flows, from a design perspective. Specifically, we have used a formalism where the coupling between (local) network structure and the conductance of individual links can be tuned to get the best results for a given objective. This approach has been inspired by the observation of correlations between link `weights' and local topology in various man-made and natural systems \cite{Barrat04,WeightNetEivind05}. Whether these systems are tuned for optimal throughput or robustness is an open question, although there have been some indications of biological systems operating in a near-optimal regime \cite{Smith1978,Kauffman1989}. Analogous weighting schemes have been proven to be efficient for optimizing flow in RW \cite{KornissweightsPRE,HuEPLbandwidth07,LingPhysLettA10} and shortest path routing \cite{YangPRE09,LingPhysLettA10,KornissBookCh}, and also for optimizing spreading \cite{Yang_PRE2008}.

The goal of this paper has been to systematically consider the optimization of throughput for distributed flows under various cases. We find that in uncorrelated scale-free and ER networks,
for identical flows between all sources and sinks and identical node capacities, each set to unity, hub-avoiding flows provide optimal throughput in the node-limited case, while weakly hub-biased flows do the same in the edge-limited case. The same trend is observed qualitatively for the Internet \cite{InternetData06}, 
and also when the flows are heterogeneous, i.e., when current between each source-target pair scales as the product of their degrees.  Interestingly though, if the capacities of nodes are also distributed heterogeneously (still conserving the total capacity), the optimal throughput occurs for a hub-biased flow. As mentioned earlier, this has possible implications for the case of airline transportation network where a similar hub-biased flow is empirically observed. Furthermore, flows on the links of metabolic networks have also been observed to be hub-biased, although a complete analogy to our distributed flow model there is unclear. Although similar trends favoring hub-avoiding flows and weakly hub-biased flows for node and edge-limited cases, respectively, are found for shortest path flows, for large systems, perhaps not surprisingly, distributed flows are more efficient at minimizing the maximal load on the network, and thus providing greater throughput. Finally, we studied how the robustness of a network to cascading failures could be optimized within the parameters of our model. Our results indicate that for small amounts of excess capacity, networks with heterogeneous degree distributions are maximally robust in the regime of hub-avoiding flows. 

\begin{acknowledgments}
We thank R. Huang for discussions in the early phase of this work.
This work was supported in part by DTRA Award No. HDTRA1-09-1-0049
and by the Army Research Laboratory under Cooperative Agreement Number W911NF-09-2-0053.
The views and conclusions contained in this document are those of the authors and should not be interpreted as representing the official policies, either expressed or implied, of the Army Research Laboratory or the U.S. Government.  The U.S. Government is authorized to reproduce and distribute reprints for Government purposes notwithstanding any copyright notation here on.
\end{acknowledgments}


\begin{thebibliography}{58}%
\makeatletter
\providecommand \@ifxundefined [1]{%
 \@ifx{#1\undefined}
}%
\providecommand \@ifnum [1]{%
 \ifnum #1\expandafter \@firstoftwo
 \else \expandafter \@secondoftwo
 \fi
}%
\providecommand \@ifx [1]{%
 \ifx #1\expandafter \@firstoftwo
 \else \expandafter \@secondoftwo
 \fi
}%
\providecommand \natexlab [1]{#1}%
\providecommand \enquote  [1]{``#1''}%
\providecommand \bibnamefont  [1]{#1}%
\providecommand \bibfnamefont [1]{#1}%
\providecommand \citenamefont [1]{#1}%
\providecommand \href@noop [0]{\@secondoftwo}%
\providecommand \href [0]{\begingroup \@sanitize@url \@href}%
\providecommand \@href[1]{\@@startlink{#1}\@@href}%
\providecommand \@@href[1]{\endgroup#1\@@endlink}%
\providecommand \@sanitize@url [0]{\catcode `\\12\catcode `\$12\catcode
  `\&12\catcode `\#12\catcode `\^12\catcode `\_12\catcode `\%12\relax}%
\providecommand \@@startlink[1]{}%
\providecommand \@@endlink[0]{}%
\providecommand \url  [0]{\begingroup\@sanitize@url \@url }%
\providecommand \@url [1]{\endgroup\@href {#1}{\urlprefix }}%
\providecommand \urlprefix  [0]{URL }%
\providecommand \Eprint [0]{\href }%
\providecommand \doibase [0]{http://dx.doi.org/}%
\providecommand \selectlanguage [0]{\@gobble}%
\providecommand \bibinfo  [0]{\@secondoftwo}%
\providecommand \bibfield  [0]{\@secondoftwo}%
\providecommand \translation [1]{[#1]}%
\providecommand \BibitemOpen [0]{}%
\providecommand \bibitemStop [0]{}%
\providecommand \bibitemNoStop [0]{.\EOS\space}%
\providecommand \EOS [0]{\spacefactor3000\relax}%
\providecommand \BibitemShut  [1]{\csname bibitem#1\endcsname}%
\let\auto@bib@innerbib\@empty
\bibitem [{\citenamefont {Guimer\`{a}}\ \emph {et~al.}(2002)\citenamefont
  {Guimer\`{a}}, \citenamefont {D\'iaz-Guilera}, \citenamefont {Vega-Redondo},
  \citenamefont {Cabrales},\ and\ \citenamefont {Arenas}}]{Guimera_PRL02}%
  \BibitemOpen
  \bibfield  {author} {\bibinfo {author} {\bibfnamefont {R.}~\bibnamefont
  {Guimer\`{a}}}, \bibinfo {author} {\bibfnamefont {A.}~\bibnamefont
  {D\'iaz-Guilera}}, \bibinfo {author} {\bibfnamefont {F.}~\bibnamefont
  {Vega-Redondo}}, \bibinfo {author} {\bibfnamefont {A.}~\bibnamefont
  {Cabrales}}, \ and\ \bibinfo {author} {\bibfnamefont {A.}~\bibnamefont
  {Arenas}},\ }\href {http://prl.aps.org/abstract/PRL/v89/i24/e248701}
  {\bibfield  {journal} {\bibinfo  {journal} {Phys. Rev. Lett}\ }\textbf
  {\bibinfo {volume} {89}},\ \bibinfo {pages} {248701} (\bibinfo {year}
  {2002})}\BibitemShut {NoStop}%
\bibitem [{\citenamefont {Yan}\ \emph {et~al.}(2006)\citenamefont {Yan},
  \citenamefont {Zhou}, \citenamefont {Hu}, \citenamefont {Fu},\ and\
  \citenamefont {Wang}}]{YanPre06}%
  \BibitemOpen
  \bibfield  {author} {\bibinfo {author} {\bibfnamefont {G.}~\bibnamefont
  {Yan}}, \bibinfo {author} {\bibfnamefont {T.}~\bibnamefont {Zhou}}, \bibinfo
  {author} {\bibfnamefont {B.}~\bibnamefont {Hu}}, \bibinfo {author}
  {\bibfnamefont {Z.-Q.}\ \bibnamefont {Fu}}, \ and\ \bibinfo {author}
  {\bibfnamefont {B.-H.}\ \bibnamefont {Wang}},\ }\href
  {http://pre.aps.org/abstract/PRE/v73/i4/e046108} {\bibfield  {journal}
  {\bibinfo  {journal} {Phys. Rev. E}\ }\textbf {\bibinfo {volume} {73}},\
  \bibinfo {pages} {046108} (\bibinfo {year} {2006})}\BibitemShut {NoStop}%
\bibitem [{\citenamefont {L\'{o}pez}\ \emph {et~al.}(2005)\citenamefont
  {L\'{o}pez}, \citenamefont {Buldyrev}, \citenamefont {Havlin},\ and\
  \citenamefont {Stanley}}]{LopezTransportPRL05}%
  \BibitemOpen
  \bibfield  {author} {\bibinfo {author} {\bibfnamefont {E.}~\bibnamefont
  {L\'{o}pez}}, \bibinfo {author} {\bibfnamefont {S.~V.}\ \bibnamefont
  {Buldyrev}}, \bibinfo {author} {\bibfnamefont {S.}~\bibnamefont {Havlin}}, \
  and\ \bibinfo {author} {\bibfnamefont {H.~E.}\ \bibnamefont {Stanley}},\
  }\href {http://prl.aps.org/abstract/PRL/v94/i24/e248701} {\bibfield
  {journal} {\bibinfo  {journal} {Phys. Rev. Lett}\ }\textbf {\bibinfo {volume}
  {94}},\ \bibinfo {pages} {248701} (\bibinfo {year} {2005})}\BibitemShut
  {NoStop}%
\bibitem [{\citenamefont {Wang}\ \emph {et~al.}(2006)\citenamefont {Wang},
  \citenamefont {Wang}, \citenamefont {Yin}, \citenamefont {Xie},\ and\
  \citenamefont {Zhou}}]{WangPre06}%
  \BibitemOpen
  \bibfield  {author} {\bibinfo {author} {\bibfnamefont {W.-X.}\ \bibnamefont
  {Wang}}, \bibinfo {author} {\bibfnamefont {B.-H.}\ \bibnamefont {Wang}},
  \bibinfo {author} {\bibfnamefont {C.-Y.}\ \bibnamefont {Yin}}, \bibinfo
  {author} {\bibfnamefont {Y.-B.}\ \bibnamefont {Xie}}, \ and\ \bibinfo
  {author} {\bibfnamefont {T.}~\bibnamefont {Zhou}},\ }\href
  {http://pre.aps.org/abstract/PRE/v73/i2/e026111} {\bibfield  {journal}
  {\bibinfo  {journal} {Phys. Rev. E}\ }\textbf {\bibinfo {volume} {73}},\
  \bibinfo {pages} {026111} (\bibinfo {year} {2006})}\BibitemShut {NoStop}%
\bibitem [{\citenamefont {Yang}\ \emph {et~al.}(2009)\citenamefont {Yang},
  \citenamefont {Wang}, \citenamefont {Lai},\ and\ \citenamefont
  {Chen}}]{YangPRE09}%
  \BibitemOpen
  \bibfield  {author} {\bibinfo {author} {\bibfnamefont {R.}~\bibnamefont
  {Yang}}, \bibinfo {author} {\bibfnamefont {W.-X.}\ \bibnamefont {Wang}},
  \bibinfo {author} {\bibfnamefont {Y.-C.}\ \bibnamefont {Lai}}, \ and\
  \bibinfo {author} {\bibfnamefont {G.}~\bibnamefont {Chen}},\ }\href
  {http://pre.aps.org/abstract/PRE/v79/i2/e026112} {\bibfield  {journal}
  {\bibinfo  {journal} {Phys. Rev. E}\ }\textbf {\bibinfo {volume} {79}},\
  \bibinfo {pages} {026112} (\bibinfo {year} {2009})}\BibitemShut {NoStop}%
\bibitem [{\citenamefont {Katifori}\ \emph {et~al.}(2010)\citenamefont
  {Katifori}, \citenamefont {Sz\"{o}ll\H{o}si},\ and\ \citenamefont
  {Magnasco}}]{Leafprl10}%
  \BibitemOpen
  \bibfield  {author} {\bibinfo {author} {\bibfnamefont {E.}~\bibnamefont
  {Katifori}}, \bibinfo {author} {\bibfnamefont {G.~J.}\ \bibnamefont
  {Sz\"{o}ll\H{o}si}}, \ and\ \bibinfo {author} {\bibfnamefont {M.~O.}\
  \bibnamefont {Magnasco}},\ }\href
  {http://prl.aps.org/abstract/PRL/v104/i4/e048704} {\bibfield  {journal}
  {\bibinfo  {journal} {Phys. Rev. Lett}\ }\textbf {\bibinfo {volume} {104}},\
  \bibinfo {pages} {048704} (\bibinfo {year} {2010})}\BibitemShut {NoStop}%
\bibitem [{\citenamefont {Sreenivasan}\ \emph {et~al.}(2007)\citenamefont
  {Sreenivasan}, \citenamefont {Cohen}, \citenamefont {L\'opez}, \citenamefont
  {Toroczkai},\ and\ \citenamefont {Stanley}}]{SameetBottleneck07}%
  \BibitemOpen
  \bibfield  {author} {\bibinfo {author} {\bibfnamefont {S.}~\bibnamefont
  {Sreenivasan}}, \bibinfo {author} {\bibfnamefont {R.}~\bibnamefont {Cohen}},
  \bibinfo {author} {\bibfnamefont {E.}~\bibnamefont {L\'opez}}, \bibinfo
  {author} {\bibfnamefont {Z.}~\bibnamefont {Toroczkai}}, \ and\ \bibinfo
  {author} {\bibfnamefont {H.~E.}\ \bibnamefont {Stanley}},\ }\href
  {http://pre.aps.org/abstract/PRE/v75/i3/e036105} {\bibfield  {journal}
  {\bibinfo  {journal} {Phys. Rev. E}\ }\textbf {\bibinfo {volume} {75}},\
  \bibinfo {pages} {036105} (\bibinfo {year} {2007})}\BibitemShut {NoStop}%
\bibitem [{\citenamefont {Danila}\ \emph
  {et~al.}(2006{\natexlab{a}})\citenamefont {Danila}, \citenamefont {Yu},
  \citenamefont {Marsh},\ and\ \citenamefont {Bassler}}]{DanilaPRE06}%
  \BibitemOpen
  \bibfield  {author} {\bibinfo {author} {\bibfnamefont {B.}~\bibnamefont
  {Danila}}, \bibinfo {author} {\bibfnamefont {Y.}~\bibnamefont {Yu}}, \bibinfo
  {author} {\bibfnamefont {J.~A.}\ \bibnamefont {Marsh}}, \ and\ \bibinfo
  {author} {\bibfnamefont {K.~E.}\ \bibnamefont {Bassler}},\ }\href
  {http://pre.aps.org/abstract/PRE/v74/i4/e046106} {\bibfield  {journal}
  {\bibinfo  {journal} {Phys. Rev. E}\ }\textbf {\bibinfo {volume} {74}},\
  \bibinfo {pages} {046106} (\bibinfo {year} {2006}{\natexlab{a}})}\BibitemShut
  {NoStop}%
\bibitem [{\citenamefont {Chandra}\ \emph {et~al.}(1997)\citenamefont
  {Chandra}, \citenamefont {Raghavan}, \citenamefont {Ruzzo}, \citenamefont
  {Smolensky},\ and\ \citenamefont {Tiwari}}]{Chandra}%
  \BibitemOpen
  \bibfield  {author} {\bibinfo {author} {\bibfnamefont {A.~K.}\ \bibnamefont
  {Chandra}}, \bibinfo {author} {\bibfnamefont {P.}~\bibnamefont {Raghavan}},
  \bibinfo {author} {\bibfnamefont {W.~L.}\ \bibnamefont {Ruzzo}}, \bibinfo
  {author} {\bibfnamefont {R.}~\bibnamefont {Smolensky}}, \ and\ \bibinfo
  {author} {\bibfnamefont {P.}~\bibnamefont {Tiwari}},\ }\href
  {http://dl.acm.org/citation.cfm?id=73062} {\bibfield  {journal} {\bibinfo
  {journal} {Comput. Complex.}\ }\textbf {\bibinfo {volume} {6}},\ \bibinfo
  {pages} {312} (\bibinfo {year} {1997})}\BibitemShut {NoStop}%
\bibitem [{\citenamefont {Aldous}\ and\ \citenamefont {Fill}(tion)}]{Aldous}%
  \BibitemOpen
  \bibfield  {author} {\bibinfo {author} {\bibfnamefont {D.}~\bibnamefont
  {Aldous}}\ and\ \bibinfo {author} {\bibfnamefont {J.}~\bibnamefont {Fill}},\
  }\href {http://www.stat.berkeley.edu/~aldous/RWG/book.html} {\enquote
  {\bibinfo {title} {Reversible {M}arkov chains and random walks on graphs},}\
  } (\bibinfo {year} {{in preparation}})\BibitemShut {NoStop}%
\bibitem [{\citenamefont {Korniss}(2007)}]{KornissweightsPRE}%
  \BibitemOpen
  \bibfield  {author} {\bibinfo {author} {\bibfnamefont {G.}~\bibnamefont
  {Korniss}},\ }\href {http://pre.aps.org/abstract/PRE/v75/i5/e051121}
  {\bibfield  {journal} {\bibinfo  {journal} {Phys. Rev. E}\ }\textbf {\bibinfo
  {volume} {75}},\ \bibinfo {pages} {051121} (\bibinfo {year}
  {2007})}\BibitemShut {NoStop}%
\bibitem [{\citenamefont {Hu}\ \emph {et~al.}(2007)\citenamefont {Hu},
  \citenamefont {Wang}, \citenamefont {Jiang}, \citenamefont {Wu},\ and\
  \citenamefont {Wu}}]{HuEPLbandwidth07}%
  \BibitemOpen
  \bibfield  {author} {\bibinfo {author} {\bibfnamefont {M.-B.}\ \bibnamefont
  {Hu}}, \bibinfo {author} {\bibfnamefont {W.-X.}\ \bibnamefont {Wang}},
  \bibinfo {author} {\bibfnamefont {R.}~\bibnamefont {Jiang}}, \bibinfo
  {author} {\bibfnamefont {Q.-S.}\ \bibnamefont {Wu}}, \ and\ \bibinfo {author}
  {\bibfnamefont {Y.-H.}\ \bibnamefont {Wu}},\ }\href
  {http://iopscience.iop.org/0295-5075/79/1/14003} {\bibfield  {journal}
  {\bibinfo  {journal} {Europhys. Lett.}\ }\textbf {\bibinfo {volume} {79}},\
  \bibinfo {pages} {14003} (\bibinfo {year} {2007})}\BibitemShut {NoStop}%
\bibitem [{\citenamefont {Ling}\ \emph {et~al.}(2010)\citenamefont {Ling},
  \citenamefont {Hu}, \citenamefont {Du}, \citenamefont {Jiang}, \citenamefont
  {Wu},\ and\ \citenamefont {Wu}}]{LingPhysLettA10}%
  \BibitemOpen
  \bibfield  {author} {\bibinfo {author} {\bibfnamefont {X.}~\bibnamefont
  {Ling}}, \bibinfo {author} {\bibfnamefont {M.-B.}\ \bibnamefont {Hu}},
  \bibinfo {author} {\bibfnamefont {W.-B.}\ \bibnamefont {Du}}, \bibinfo
  {author} {\bibfnamefont {R.}~\bibnamefont {Jiang}}, \bibinfo {author}
  {\bibfnamefont {Y.-H.}\ \bibnamefont {Wu}}, \ and\ \bibinfo {author}
  {\bibfnamefont {Q.-S.}\ \bibnamefont {Wu}},\ }\href
  {http://www.sciencedirect.com/science/article/pii/S0375960110013654}
  {\bibfield  {journal} {\bibinfo  {journal} {Phys. Lett. A}\ }\textbf
  {\bibinfo {volume} {374}},\ \bibinfo {pages} {4825} (\bibinfo {year}
  {2010})}\BibitemShut {NoStop}%
\bibitem [{\citenamefont {Danila}\ \emph
  {et~al.}(2006{\natexlab{b}})\citenamefont {Danila}, \citenamefont {Yu},
  \citenamefont {Earl}, \citenamefont {Marsh}, \citenamefont {Toroczkai},\ and\
  \citenamefont {Bassler}}]{Danila_RW_PRE2006}%
  \BibitemOpen
  \bibfield  {author} {\bibinfo {author} {\bibfnamefont {B.}~\bibnamefont
  {Danila}}, \bibinfo {author} {\bibfnamefont {Y.}~\bibnamefont {Yu}}, \bibinfo
  {author} {\bibfnamefont {S.}~\bibnamefont {Earl}}, \bibinfo {author}
  {\bibfnamefont {J.~A.}\ \bibnamefont {Marsh}}, \bibinfo {author}
  {\bibfnamefont {Z.}~\bibnamefont {Toroczkai}}, \ and\ \bibinfo {author}
  {\bibfnamefont {K.~E.}\ \bibnamefont {Bassler}},\ }\href
  {http://pre.aps.org/abstract/PRE/v74/i4/e046114} {\bibfield  {journal}
  {\bibinfo  {journal} {Phys. Rev. E}\ }\textbf {\bibinfo {volume} {74}},\
  \bibinfo {pages} {046114} (\bibinfo {year} {2006}{\natexlab{b}})}\BibitemShut
  {NoStop}%
\bibitem [{\citenamefont {Asztalos}\ and\ \citenamefont
  {Toroczkai}(2010)}]{Asztalos_EPL2010}%
  \BibitemOpen
  \bibfield  {author} {\bibinfo {author} {\bibfnamefont {A.}~\bibnamefont
  {Asztalos}}\ and\ \bibinfo {author} {\bibfnamefont {Z.}~\bibnamefont
  {Toroczkai}},\ }\href {http://iopscience.iop.org/0295-5075/92/5/50008}
  {\bibfield  {journal} {\bibinfo  {journal} {Europhys. Lett.}\ }\textbf
  {\bibinfo {volume} {92}},\ \bibinfo {pages} {50008} (\bibinfo {year}
  {2010})}\BibitemShut {NoStop}%
\bibitem [{\citenamefont {{J.S. Andrade, Jr.}}\ \emph
  {et~al.}(2005)\citenamefont {{J.S. Andrade, Jr.}}, \citenamefont {Herrmann},
  \citenamefont {Andrade},\ and\ \citenamefont {da~Silva}}]{Andrade05}%
  \BibitemOpen
  \bibfield  {author} {\bibinfo {author} {\bibnamefont {{J.S. Andrade, Jr.}}},
  \bibinfo {author} {\bibfnamefont {H.~J.}\ \bibnamefont {Herrmann}}, \bibinfo
  {author} {\bibfnamefont {R.~F.~S.}\ \bibnamefont {Andrade}}, \ and\ \bibinfo
  {author} {\bibfnamefont {L.~R.}\ \bibnamefont {da~Silva}},\ }\href
  {http://prl.aps.org/abstract/PRL/v94/i1/e018702} {\bibfield  {journal}
  {\bibinfo  {journal} {Phys. Rev. Lett}\ }\textbf {\bibinfo {volume} {94}},\
  \bibinfo {pages} {018702} (\bibinfo {year} {2005})}\BibitemShut {NoStop}%
\bibitem [{\citenamefont {Lee}\ and\ \citenamefont
  {Rieger}(2006)}]{DsLee_epl06}%
  \BibitemOpen
  \bibfield  {author} {\bibinfo {author} {\bibfnamefont {D.-S.}\ \bibnamefont
  {Lee}}\ and\ \bibinfo {author} {\bibfnamefont {H.}~\bibnamefont {Rieger}},\
  }\href {http://iopscience.iop.org/0295-5075/73/3/471} {\bibfield  {journal}
  {\bibinfo  {journal} {Europhys. Lett.}\ }\textbf {\bibinfo {volume} {73}},\
  \bibinfo {pages} {471} (\bibinfo {year} {2006})}\BibitemShut {NoStop}%
\bibitem [{\citenamefont {Korniss}\ \emph {et~al.}(2006)\citenamefont
  {Korniss}, \citenamefont {Hastings}, \citenamefont {Bassler}, \citenamefont
  {Berryman}, \citenamefont {Kozma},\ and\ \citenamefont
  {Abbott}}]{KornissSWPhysLettA06}%
  \BibitemOpen
  \bibfield  {author} {\bibinfo {author} {\bibfnamefont {G.}~\bibnamefont
  {Korniss}}, \bibinfo {author} {\bibfnamefont {M.~B.}\ \bibnamefont
  {Hastings}}, \bibinfo {author} {\bibfnamefont {K.~E.}\ \bibnamefont
  {Bassler}}, \bibinfo {author} {\bibfnamefont {M.~J.}\ \bibnamefont
  {Berryman}}, \bibinfo {author} {\bibfnamefont {B.}~\bibnamefont {Kozma}}, \
  and\ \bibinfo {author} {\bibfnamefont {D.}~\bibnamefont {Abbott}},\ }\href
  {http://www.sciencedirect.com/science/article/pii/S0375960105016026}
  {\bibfield  {journal} {\bibinfo  {journal} {Phys. Lett. A}\ }\textbf
  {\bibinfo {volume} {350}},\ \bibinfo {pages} {324} (\bibinfo {year}
  {2006})}\BibitemShut {NoStop}%
\bibitem [{\citenamefont {Prusinkiewicz}\ \emph {et~al.}(2007)\citenamefont
  {Prusinkiewicz}, \citenamefont {Allen}, \citenamefont {Escobar-Guti\'errez},\
  and\ \citenamefont {Dejong}}]{Plant07}%
  \BibitemOpen
  \bibfield  {author} {\bibinfo {author} {\bibfnamefont {P.}~\bibnamefont
  {Prusinkiewicz}}, \bibinfo {author} {\bibfnamefont {M.}~\bibnamefont
  {Allen}}, \bibinfo {author} {\bibfnamefont {A.}~\bibnamefont
  {Escobar-Guti\'errez}}, \ and\ \bibinfo {author} {\bibfnamefont {T.~M.}\
  \bibnamefont {Dejong}},\ }\href
  {http://www.springer.com/life+sciences/plant+sciences/book/978-1-4020-6033-5}
  {\emph {\bibinfo {title} {Functional-Structural Plant Modelling in Crop
  Production}}},\ edited by\ \bibinfo {editor} {\bibfnamefont {J.}~\bibnamefont
  {Vos}}, \bibinfo {editor} {\bibfnamefont {L.~F.~M.}\ \bibnamefont
  {Marcelis}}, \bibinfo {editor} {\bibfnamefont {P.~H.~B.}\ \bibnamefont
  {de~Visser}}, \bibinfo {editor} {\bibfnamefont {P.~C.}\ \bibnamefont
  {Struik}}, \ and\ \bibinfo {editor} {\bibfnamefont {J.~B.}\ \bibnamefont
  {Evers}}\ (\bibinfo  {publisher} {Springer},\ \bibinfo {year} {2007})\
  Chap.~\bibinfo {chapter} {11}, pp.\ \bibinfo {pages} {123--137}\BibitemShut
  {NoStop}%
\bibitem [{\citenamefont {Fatt}(1956)}]{PorousFatt56}%
  \BibitemOpen
  \bibfield  {author} {\bibinfo {author} {\bibfnamefont {I.}~\bibnamefont
  {Fatt}},\ }\href {http://www.aimehq.org/search/docs/Volume%20207/207-34.pdf}
  {\bibfield  {journal} {\bibinfo  {journal} {Trans. AIME}\ }\textbf {\bibinfo
  {volume} {207}},\ \bibinfo {pages} {144} (\bibinfo {year}
  {1956})}\BibitemShut {NoStop}%
\bibitem [{\citenamefont {Kirkpatrick}(1971)}]{KirkpatrickPRL71}%
  \BibitemOpen
  \bibfield  {author} {\bibinfo {author} {\bibfnamefont {S.}~\bibnamefont
  {Kirkpatrick}},\ }\href {http://prl.aps.org/abstract/PRL/v27/i25/p1722_1}
  {\bibfield  {journal} {\bibinfo  {journal} {Phys. Rev. Lett}\ }\textbf
  {\bibinfo {volume} {27}},\ \bibinfo {pages} {1722} (\bibinfo {year}
  {1971})}\BibitemShut {NoStop}%
\bibitem [{\citenamefont {Newman}\ and\ \citenamefont
  {Girvan}(2004)}]{NewmanGirvan04}%
  \BibitemOpen
  \bibfield  {author} {\bibinfo {author} {\bibfnamefont {M.~E.~J.}\
  \bibnamefont {Newman}}\ and\ \bibinfo {author} {\bibfnamefont
  {M.}~\bibnamefont {Girvan}},\ }\href
  {http://pre.aps.org/abstract/PRE/v69/i2/e026113} {\bibfield  {journal}
  {\bibinfo  {journal} {Phys. Rev. E}\ }\textbf {\bibinfo {volume} {69}},\
  \bibinfo {pages} {026113} (\bibinfo {year} {2004})}\BibitemShut {NoStop}%
\bibitem [{\citenamefont {Kaul}\ \emph {et~al.}(2009)\citenamefont {Kaul},
  \citenamefont {Yun},\ and\ \citenamefont {{S.-G. Kim}}}]{Kaul_ACM2009}%
  \BibitemOpen
  \bibfield  {author} {\bibinfo {author} {\bibfnamefont {R.}~\bibnamefont
  {Kaul}}, \bibinfo {author} {\bibfnamefont {Y.}~\bibnamefont {Yun}}, \ and\
  \bibinfo {author} {\bibnamefont {{S.-G. Kim}}},\ }\href
  {http://dl.acm.org/citation.cfm?doid=1536616.1536649} {\bibfield  {journal}
  {\bibinfo  {journal} {Commun. ACM}\ }\textbf {\bibinfo {volume} {52}},\
  \bibinfo {pages} {132} (\bibinfo {year} {2009})}\BibitemShut {NoStop}%
\bibitem [{\citenamefont {Doyle}\ and\ \citenamefont
  {Snell}(1984)}]{DoyleSnell84}%
  \BibitemOpen
  \bibfield  {author} {\bibinfo {author} {\bibfnamefont {P.~G.}\ \bibnamefont
  {Doyle}}\ and\ \bibinfo {author} {\bibfnamefont {J.~L.}\ \bibnamefont
  {Snell}},\ }\href {http://www.stanford.edu/class/msande337/notes/walks.pdf}
  {\bibfield  {journal} {\bibinfo  {journal} {The Mathematical Association of
  America}\ }\textbf {\bibinfo {volume} {22}} (\bibinfo {year}
  {1984})}\BibitemShut {NoStop}%
\bibitem [{\citenamefont {Tetali}(1991)}]{Tetali_JTP1991}%
  \BibitemOpen
  \bibfield  {author} {\bibinfo {author} {\bibfnamefont {P.}~\bibnamefont
  {Tetali}},\ }\href {http://www.springerlink.com/content/mqx05041478g6773/}
  {\bibfield  {journal} {\bibinfo  {journal} {J. Theor. Probab.}\ }\textbf
  {\bibinfo {volume} {4}},\ \bibinfo {pages} {101} (\bibinfo {year}
  {1991})}\BibitemShut {NoStop}%
\bibitem [{\citenamefont {Redner}()}]{Redner07}%
  \BibitemOpen
  \bibfield  {author} {\bibinfo {author} {\bibfnamefont {S.}~\bibnamefont
  {Redner}},\ }\href {http://arxiv.org/abs/0710.1105} {\bibinfo  {journal}
  {arXiv:0710.1105v1}\ }\BibitemShut {NoStop}%
\bibitem [{\citenamefont {Ghosh}\ \emph {et~al.}(2008)\citenamefont {Ghosh},
  \citenamefont {Boyd},\ and\ \citenamefont {Saberi}}]{GBS_SIAM2008}%
  \BibitemOpen
\bibfield  {journal} {  }\bibfield  {author} {\bibinfo {author} {\bibfnamefont
  {A.}~\bibnamefont {Ghosh}}, \bibinfo {author} {\bibfnamefont
  {S.}~\bibnamefont {Boyd}}, \ and\ \bibinfo {author} {\bibfnamefont
  {A.}~\bibnamefont {Saberi}},\ }\href
  {http://dl.acm.org/citation.cfm?id=1350629} {\bibfield  {journal} {\bibinfo
  {journal} {SIAM Rev.}\ }\textbf {\bibinfo {volume} {50}},\ \bibinfo {pages}
  {37} (\bibinfo {year} {2008})}\BibitemShut {NoStop}%
\bibitem [{\citenamefont {{W. Ellens et al.}}(2011)}]{Ellens_LAA2011}%
  \BibitemOpen
  \bibfield  {author} {\bibinfo {author} {\bibnamefont {{W. Ellens et al.}}},\
  }\href
  {http://www.nas.ewi.tudelft.nl/people/Piet/papers/LAA_2011_EffectiveResistance.pdf}
  {\bibfield  {journal} {\bibinfo  {journal} {Linear Algebra Appl.}\ }\textbf
  {\bibinfo {volume} {435}},\ \bibinfo {pages} {2491} (\bibinfo {year}
  {2011})}\BibitemShut {NoStop}%
\bibitem [{\citenamefont {Tizghadam}\ and\ \citenamefont
  {Leon-Garcia}(2008)}]{TLG_GTC2008}%
  \BibitemOpen
  \bibfield  {author} {\bibinfo {author} {\bibfnamefont {A.}~\bibnamefont
  {Tizghadam}}\ and\ \bibinfo {author} {\bibfnamefont {A.}~\bibnamefont
  {Leon-Garcia}},\ }in\ \href
  {http://ieeexplore.ieee.org/Xplore/login.jsp?url=http%3A%2F%2Fieeexplore.ieee.org%2Fiel5%2F4697774%2F4697775%2F04698231.pdf%3Farnumber%3D4698231&authDecision=-203}
  {\emph {\bibinfo {booktitle} {Proc. of IEEE Global Telecomm. Conf.}}}\
  (\bibinfo {year} {2008})\ pp.\ \bibinfo {pages} {1--6}\BibitemShut {NoStop}%
\bibitem [{\citenamefont {Tizghadam}\ and\ \citenamefont
  {Leon-Garcia}(2010)}]{TLG_IEEE2010}%
  \BibitemOpen
  \bibfield  {author} {\bibinfo {author} {\bibfnamefont {A.}~\bibnamefont
  {Tizghadam}}\ and\ \bibinfo {author} {\bibfnamefont {A.}~\bibnamefont
  {Leon-Garcia}},\ }\href
  {http://ieeexplore.ieee.org/xpl/freeabs_all.jsp?arnumber=5634437} {\bibfield
  {journal} {\bibinfo  {journal} {IEEE Network}\ }\textbf {\bibinfo {volume}
  {24}},\ \bibinfo {pages} {10} (\bibinfo {year} {2010})}\BibitemShut {NoStop}%
\bibitem [{\citenamefont {Wu}\ \emph {et~al.}(2006)\citenamefont {Wu},
  \citenamefont {Braunstein}, \citenamefont {Colizza}, \citenamefont {Cohen},
  \citenamefont {Havlin},\ and\ \citenamefont {Stanley}}]{Colizza06}%
  \BibitemOpen
  \bibfield  {author} {\bibinfo {author} {\bibfnamefont {Z.}~\bibnamefont
  {Wu}}, \bibinfo {author} {\bibfnamefont {L.~A.}\ \bibnamefont {Braunstein}},
  \bibinfo {author} {\bibfnamefont {V.}~\bibnamefont {Colizza}}, \bibinfo
  {author} {\bibfnamefont {R.}~\bibnamefont {Cohen}}, \bibinfo {author}
  {\bibfnamefont {S.}~\bibnamefont {Havlin}}, \ and\ \bibinfo {author}
  {\bibfnamefont {H.~E.}\ \bibnamefont {Stanley}},\ }\href
  {http://pre.aps.org/abstract/PRE/v74/i5/e056104} {\bibfield  {journal}
  {\bibinfo  {journal} {Phys. Rev. E}\ }\textbf {\bibinfo {volume} {74}},\
  \bibinfo {pages} {056104} (\bibinfo {year} {2006})}\BibitemShut {NoStop}%
\bibitem [{\citenamefont {Baronchelli}\ and\ \citenamefont
  {Pastor-{S}atorras}(2010)}]{Baron10}%
  \BibitemOpen
  \bibfield  {author} {\bibinfo {author} {\bibfnamefont {A.}~\bibnamefont
  {Baronchelli}}\ and\ \bibinfo {author} {\bibfnamefont {R.}~\bibnamefont
  {Pastor-{S}atorras}},\ }\href
  {http://pre.aps.org/abstract/PRE/v82/i1/e011111} {\bibfield  {journal}
  {\bibinfo  {journal} {Phys. Rev. E}\ }\textbf {\bibinfo {volume} {82}},\
  \bibinfo {pages} {011111} (\bibinfo {year} {2010})}\BibitemShut {NoStop}%
\bibitem [{\citenamefont {Barrat}\ \emph {et~al.}(2004)\citenamefont {Barrat},
  \citenamefont {Barth\'{e}lemy}, \citenamefont {Pastor-{S}atorras},\ and\
  \citenamefont {Vespignani}}]{Barrat04}%
  \BibitemOpen
  \bibfield  {author} {\bibinfo {author} {\bibfnamefont {A.}~\bibnamefont
  {Barrat}}, \bibinfo {author} {\bibfnamefont {M.}~\bibnamefont
  {Barth\'{e}lemy}}, \bibinfo {author} {\bibfnamefont {R.}~\bibnamefont
  {Pastor-{S}atorras}}, \ and\ \bibinfo {author} {\bibfnamefont
  {A.}~\bibnamefont {Vespignani}},\ }\href
  {http://www.pnas.org/content/101/11/3747.abstract?sid=6edbcb29-3f66-4a9f-82e8-217828bbada4}
  {\bibfield  {journal} {\bibinfo  {journal} {Proc. Natl. Acad. Sci. USA}\
  }\textbf {\bibinfo {volume} {101}},\ \bibinfo {pages} {3747} (\bibinfo {year}
  {2004})}\BibitemShut {NoStop}%
\bibitem [{\citenamefont {MacDonald}\ \emph {et~al.}(2005)\citenamefont
  {MacDonald}, \citenamefont {Almaas},\ and\ \citenamefont
  {Barab\'asi}}]{WeightNetEivind05}%
  \BibitemOpen
  \bibfield  {author} {\bibinfo {author} {\bibfnamefont {P.~J.}\ \bibnamefont
  {MacDonald}}, \bibinfo {author} {\bibfnamefont {E.}~\bibnamefont {Almaas}}, \
  and\ \bibinfo {author} {\bibfnamefont {A.-L.}\ \bibnamefont {Barab\'asi}},\
  }\href
  {http://epljournal.edpsciences.org/index.php?option=com_article&access=standard&Itemid=129&url=/articles/epl/abs/2005/20/epl8979/epl8979.html}
  {\bibfield  {journal} {\bibinfo  {journal} {Europhys. Lett.}\ }\textbf
  {\bibinfo {volume} {72}},\ \bibinfo {pages} {308} (\bibinfo {year}
  {2005})}\BibitemShut {NoStop}%
\bibitem [{\citenamefont {Korniss}\ \emph {et~al.}(ress)\citenamefont
  {Korniss}, \citenamefont {Huang}, \citenamefont {Sreenivasan},\ and\
  \citenamefont {Szymanski}}]{KornissBookCh}%
  \BibitemOpen
  \bibfield  {author} {\bibinfo {author} {\bibfnamefont {G.}~\bibnamefont
  {Korniss}}, \bibinfo {author} {\bibfnamefont {R.}~\bibnamefont {Huang}},
  \bibinfo {author} {\bibfnamefont {S.}~\bibnamefont {Sreenivasan}}, \ and\
  \bibinfo {author} {\bibfnamefont {B.~K.}\ \bibnamefont {Szymanski}},\ }in\
  \href@noop {} {\emph {\bibinfo {booktitle} {Handbook of optimization in
  complex networks}}},\ \bibinfo {series} {Springer Optimization and Its
  Applications}, Vol.~\bibinfo {volume} {58},\ \bibinfo {editor} {edited by\
  \bibinfo {editor} {\bibfnamefont {T.}~\bibnamefont {Thai}}\ and\ \bibinfo
  {editor} {\bibfnamefont {P.}~\bibnamefont {Pardalos}}}\ (\bibinfo
  {publisher} {Springer},\ \bibinfo {year} {2011, in press})\BibitemShut
  {NoStop}%
\bibitem [{\citenamefont {Hernandez}\ \emph {et~al.}(2005)\citenamefont
  {Hernandez}, \citenamefont {Roman},\ and\ \citenamefont
  {Vidal}}]{Slepc_Hernandez}%
  \BibitemOpen
  \bibfield  {author} {\bibinfo {author} {\bibfnamefont {V.}~\bibnamefont
  {Hernandez}}, \bibinfo {author} {\bibfnamefont {J.~E.}\ \bibnamefont
  {Roman}}, \ and\ \bibinfo {author} {\bibfnamefont {V.}~\bibnamefont
  {Vidal}},\ }\href {http://dl.acm.org/citation.cfm?id=1089019} {\bibfield
  {journal} {\bibinfo  {journal} {ACM Trans. Math. Software}\ }\textbf
  {\bibinfo {volume} {31}},\ \bibinfo {pages} {351} (\bibinfo {year}
  {2005})}\BibitemShut {NoStop}%
\bibitem [{\citenamefont {Newman}(2005)}]{Newman_betw}%
  \BibitemOpen
  \bibfield  {author} {\bibinfo {author} {\bibfnamefont {M.~E.~J.}\
  \bibnamefont {Newman}},\ }\href
  {http://www.sciencedirect.com/science/article/pii/S0378873304000681}
  {\bibfield  {journal} {\bibinfo  {journal} {Social Networks}\ }\textbf
  {\bibinfo {volume} {27}},\ \bibinfo {pages} {39} (\bibinfo {year}
  {2005})}\BibitemShut {NoStop}%
\bibitem [{\citenamefont {Brandes}\ and\ \citenamefont
  {Fleischer}(2005)}]{BrandesCflow}%
  \BibitemOpen
  \bibfield  {author} {\bibinfo {author} {\bibfnamefont {U.}~\bibnamefont
  {Brandes}}\ and\ \bibinfo {author} {\bibfnamefont {D.}~\bibnamefont
  {Fleischer}},\ }in\ \href
  {http://www.springerlink.com/content/jtuxnqb4eqlmlqbj/} {\emph {\bibinfo
  {booktitle} {Lecture Notes in Computer Science}}},\ Vol.\ \bibinfo {volume}
  {3404},\ \bibinfo {editor} {edited by\ \bibinfo {editor} {\bibfnamefont
  {V.}~\bibnamefont {Diekert}}\ and\ \bibinfo {editor} {\bibfnamefont
  {B.}~\bibnamefont {Durand}}}\ (\bibinfo  {publisher} {Springer, NY},\
  \bibinfo {year} {2005})\ pp.\ \bibinfo {pages} {533--544}\BibitemShut
  {NoStop}%
\bibitem [{\citenamefont {Goh}\ \emph {et~al.}(2001)\citenamefont {Goh},
  \citenamefont {Kahng},\ and\ \citenamefont {Kim}}]{GohLoad01}%
  \BibitemOpen
  \bibfield  {author} {\bibinfo {author} {\bibfnamefont {K.-I.}\ \bibnamefont
  {Goh}}, \bibinfo {author} {\bibfnamefont {B.}~\bibnamefont {Kahng}}, \ and\
  \bibinfo {author} {\bibfnamefont {D.}~\bibnamefont {Kim}},\ }\href
  {http://prl.aps.org/abstract/PRL/v87/i27/e278701} {\bibfield  {journal}
  {\bibinfo  {journal} {Phys. Rev. Lett}\ }\textbf {\bibinfo {volume} {87}},\
  \bibinfo {pages} {278701} (\bibinfo {year} {2001})}\BibitemShut {NoStop}%
\bibitem [{\citenamefont {Goh}\ \emph {et~al.}(2003)\citenamefont {Goh},
  \citenamefont {Kahng},\ and\ \citenamefont {Kim}}]{Goh_PhysA03}%
  \BibitemOpen
  \bibfield  {author} {\bibinfo {author} {\bibfnamefont {K.-I.}\ \bibnamefont
  {Goh}}, \bibinfo {author} {\bibfnamefont {B.}~\bibnamefont {Kahng}}, \ and\
  \bibinfo {author} {\bibfnamefont {D.}~\bibnamefont {Kim}},\ }\href
  {http://www.sciencedirect.com/science/article/pii/S0378437102014073}
  {\bibfield  {journal} {\bibinfo  {journal} {Physica A}\ }\textbf {\bibinfo
  {volume} {318}},\ \bibinfo {pages} {72} (\bibinfo {year} {2003})}\BibitemShut
  {NoStop}%
\bibitem [{\citenamefont {Molloy}\ and\ \citenamefont
  {Reed}(1995)}]{MolloyR95}%
  \BibitemOpen
  \bibfield  {author} {\bibinfo {author} {\bibfnamefont {M.}~\bibnamefont
  {Molloy}}\ and\ \bibinfo {author} {\bibfnamefont {B.~A.}\ \bibnamefont
  {Reed}},\ }\href
  {http://onlinelibrary.wiley.com/doi/10.1002/rsa.3240060204/abstract}
  {\bibfield  {journal} {\bibinfo  {journal} {Random Struct. Algorithms}\
  }\textbf {\bibinfo {volume} {6}},\ \bibinfo {pages} {161} (\bibinfo {year}
  {1995})}\BibitemShut {NoStop}%
\bibitem [{\citenamefont {Bogu{\~n}{\'a}}\ \emph {et~al.}(2004)\citenamefont
  {Bogu{\~n}{\'a}}, \citenamefont {Pastor-Satorras},\ and\ \citenamefont
  {Vespignani}}]{BogunaUCM04}%
  \BibitemOpen
  \bibfield  {author} {\bibinfo {author} {\bibfnamefont {M.}~\bibnamefont
  {Bogu{\~n}{\'a}}}, \bibinfo {author} {\bibfnamefont {R.}~\bibnamefont
  {Pastor-Satorras}}, \ and\ \bibinfo {author} {\bibfnamefont {A.}~\bibnamefont
  {Vespignani}},\ }\href
  {http://www.springerlink.com/content/wpc6g51eu35fe2bq/} {\bibfield  {journal}
  {\bibinfo  {journal} {Eur. Phys. J. B}\ }\textbf {\bibinfo {volume} {38}},\
  \bibinfo {pages} {205} (\bibinfo {year} {2004})}\BibitemShut {NoStop}%
\bibitem [{\citenamefont {Huang}(2010)}]{Huang_MS2010}%
  \BibitemOpen
  \bibfield  {author} {\bibinfo {author} {\bibfnamefont {R.}~\bibnamefont
  {Huang}},\ }\emph {\bibinfo {title} {Flow Optimization in Complex
  Networks}},\ \href@noop {} {Master's thesis},\ \bibinfo  {school} {Rensselaer
  Polytechnic Institute}, \bibinfo {address} {Troy, NY} (\bibinfo {year}
  {2010})\BibitemShut {NoStop}%
\bibitem [{\citenamefont {Barab\'asi}\ and\ \citenamefont
  {Albert}(1999)}]{Science_BA99}%
  \BibitemOpen
  \bibfield  {author} {\bibinfo {author} {\bibfnamefont {A.-L.}\ \bibnamefont
  {Barab\'asi}}\ and\ \bibinfo {author} {\bibfnamefont {R.}~\bibnamefont
  {Albert}},\ }\href {http://www.sciencemag.org/content/286/5439/509}
  {\bibfield  {journal} {\bibinfo  {journal} {Science}\ }\textbf {\bibinfo
  {volume} {286}},\ \bibinfo {pages} {509} (\bibinfo {year}
  {1999})}\BibitemShut {NoStop}%
\bibitem [{\citenamefont {Erd\H{o}s}\ and\ \citenamefont
  {R\'{e}nyi}(1959)}]{ErdosRgraph}%
  \BibitemOpen
  \bibfield  {author} {\bibinfo {author} {\bibfnamefont {P.}~\bibnamefont
  {Erd\H{o}s}}\ and\ \bibinfo {author} {\bibfnamefont {A.}~\bibnamefont
  {R\'{e}nyi}},\ }\href {http://www.citeulike.org/group/3509/article/4012374}
  {\bibfield  {journal} {\bibinfo  {journal} {Public. Math.}\ }\textbf
  {\bibinfo {volume} {6}},\ \bibinfo {pages} {290} (\bibinfo {year}
  {1959})}\BibitemShut {NoStop}%
\bibitem [{Int()}]{InternetData06}%
  \BibitemOpen
  \href@noop {} {}\bibinfo {howpublished}
  {http://www-personal.umich.edu/~mejn/netdata/}\BibitemShut {NoStop}%
\bibitem [{\citenamefont {Newman}(2003)}]{MixingNewman03}%
  \BibitemOpen
  \bibfield  {author} {\bibinfo {author} {\bibfnamefont {M.~E.~J.}\
  \bibnamefont {Newman}},\ }\href
  {http://pre.aps.org/abstract/PRE/v67/i2/e026126} {\bibfield  {journal}
  {\bibinfo  {journal} {Phys. Rev. E}\ }\textbf {\bibinfo {volume} {67}},\
  \bibinfo {pages} {026126} (\bibinfo {year} {2003})}\BibitemShut {NoStop}%
\bibitem [{\citenamefont {Freeman}(1977)}]{Freeman77}%
  \BibitemOpen
  \bibfield  {author} {\bibinfo {author} {\bibfnamefont {L.~C.}\ \bibnamefont
  {Freeman}},\ }\href {http://www.jstor.org/pss/3033543} {\bibfield  {journal}
  {\bibinfo  {journal} {Sociometry}\ }\textbf {\bibinfo {volume} {40}},\
  \bibinfo {pages} {35} (\bibinfo {year} {1977})}\BibitemShut {NoStop}%
\bibitem [{\citenamefont {Brandes}(2008)}]{Brandes08}%
  \BibitemOpen
  \bibfield  {author} {\bibinfo {author} {\bibfnamefont {U.}~\bibnamefont
  {Brandes}},\ }\href
  {http://www.sciencedirect.com/science/article/pii/S0378873307000731}
  {\bibfield  {journal} {\bibinfo  {journal} {Social Networks}\ }\textbf
  {\bibinfo {volume} {30}},\ \bibinfo {pages} {136} (\bibinfo {year}
  {2008})}\BibitemShut {NoStop}%
\bibitem [{\citenamefont {Jacobson}(1988)}]{Congestion88}%
  \BibitemOpen
  \bibfield  {author} {\bibinfo {author} {\bibfnamefont {V.}~\bibnamefont
  {Jacobson}},\ }in\ \href {http://dl.acm.org/citation.cfm?id=52356} {\emph
  {\bibinfo {booktitle} {Proc. of ACM SIGCOMM '88}}}\ (\bibinfo  {publisher}
  {Stanford, CA},\ \bibinfo {year} {1988})\ pp.\ \bibinfo {pages}
  {314--329}\BibitemShut {NoStop}%
\bibitem [{\citenamefont {Sachs}(2009)}]{Sachs09}%
  \BibitemOpen
  \bibfield  {author} {\bibinfo {author} {\bibfnamefont {J.~D.}\ \bibnamefont
  {Sachs}},\ }\href
  {http://www.sciamdigital.com/index.cfm?fa=Products.ViewIssuePreview&ARTICLEID_CHAR=1D86B396-3048-8A5E-108709894EA98797}
  {\bibfield  {journal} {\bibinfo  {journal} {Scientific American}\ }\textbf
  {\bibinfo {volume} {300}},\ \bibinfo {pages} {34} (\bibinfo {year}
  {2009})}\BibitemShut {NoStop}%
\bibitem [{\citenamefont {Moreno}\ \emph {et~al.}(2002)\citenamefont {Moreno},
  \citenamefont {G\'{o}mez},\ and\ \citenamefont {Pacheco}}]{MorenoEPL02}%
  \BibitemOpen
  \bibfield  {author} {\bibinfo {author} {\bibfnamefont {Y.}~\bibnamefont
  {Moreno}}, \bibinfo {author} {\bibfnamefont {J.~B.}\ \bibnamefont
  {G\'{o}mez}}, \ and\ \bibinfo {author} {\bibfnamefont {A.~F.}\ \bibnamefont
  {Pacheco}},\ }\href {http://iopscience.iop.org/0295-5075/58/4/630} {\bibfield
   {journal} {\bibinfo  {journal} {Europhys. Lett.}\ }\textbf {\bibinfo
  {volume} {58}},\ \bibinfo {pages} {630} (\bibinfo {year} {2002})}\BibitemShut
  {NoStop}%
\bibitem [{\citenamefont {Motter}\ and\ \citenamefont
  {Lai}(2002)}]{AdilsonCasc02}%
  \BibitemOpen
  \bibfield  {author} {\bibinfo {author} {\bibfnamefont {A.~E.}\ \bibnamefont
  {Motter}}\ and\ \bibinfo {author} {\bibfnamefont {Y.-C.}\ \bibnamefont
  {Lai}},\ }\href {http://pre.aps.org/abstract/PRE/v66/i6/e065102} {\bibfield
  {journal} {\bibinfo  {journal} {Phys. Rev. E}\ }\textbf {\bibinfo {volume}
  {66}},\ \bibinfo {pages} {065102(R)} (\bibinfo {year} {2002})}\BibitemShut
  {NoStop}%
\bibitem [{\citenamefont {Holme}\ \emph {et~al.}(2002)\citenamefont {Holme},
  \citenamefont {Kim}, \citenamefont {Yoon},\ and\ \citenamefont
  {Han}}]{HolmeattackPre02}%
  \BibitemOpen
  \bibfield  {author} {\bibinfo {author} {\bibfnamefont {P.}~\bibnamefont
  {Holme}}, \bibinfo {author} {\bibfnamefont {B.~J.}\ \bibnamefont {Kim}},
  \bibinfo {author} {\bibfnamefont {C.~N.}\ \bibnamefont {Yoon}}, \ and\
  \bibinfo {author} {\bibfnamefont {S.~K.}\ \bibnamefont {Han}},\ }\href
  {http://pre.aps.org/abstract/PRE/v65/i5/e056109} {\bibfield  {journal}
  {\bibinfo  {journal} {Phys. Rev. E}\ }\textbf {\bibinfo {volume} {65}},\
  \bibinfo {pages} {056109} (\bibinfo {year} {2002})}\BibitemShut {NoStop}%
\bibitem [{\citenamefont {Crucitti}\ \emph {et~al.}(2004)\citenamefont
  {Crucitti}, \citenamefont {Latora},\ and\ \citenamefont
  {Marchiori}}]{CrucittiPre04}%
  \BibitemOpen
  \bibfield  {author} {\bibinfo {author} {\bibfnamefont {P.}~\bibnamefont
  {Crucitti}}, \bibinfo {author} {\bibfnamefont {V.}~\bibnamefont {Latora}}, \
  and\ \bibinfo {author} {\bibfnamefont {M.}~\bibnamefont {Marchiori}},\ }\href
  {http://pre.aps.org/abstract/PRE/v69/i4/e045104} {\bibfield  {journal}
  {\bibinfo  {journal} {Phys. Rev. E}\ }\textbf {\bibinfo {volume} {69}},\
  \bibinfo {pages} {045104(R)} (\bibinfo {year} {2004})}\BibitemShut {NoStop}%
\bibitem [{\citenamefont {Smith}(1978)}]{Smith1978}%
  \BibitemOpen
  \bibfield  {author} {\bibinfo {author} {\bibfnamefont {J.~M.}\ \bibnamefont
  {Smith}},\ }\href {http://www.jstor.org/stable/2096742} {\bibfield  {journal}
  {\bibinfo  {journal} {Annu. Rev. Ecol. Syst.}\ }\textbf {\bibinfo {volume}
  {9}},\ \bibinfo {pages} {pp. 31} (\bibinfo {year} {1978})}\BibitemShut
  {NoStop}%
\bibitem [{\citenamefont {Kauffman}\ and\ \citenamefont
  {Weinberger}(1989)}]{Kauffman1989}%
  \BibitemOpen
  \bibfield  {author} {\bibinfo {author} {\bibfnamefont {S.~A.}\ \bibnamefont
  {Kauffman}}\ and\ \bibinfo {author} {\bibfnamefont {E.~D.}\ \bibnamefont
  {Weinberger}},\ }\href {\doibase DOI: 10.1016/S0022-5193(89)80019-0}
  {\bibfield  {journal} {\bibinfo  {journal} {J. Theor. Probab.}\ }\textbf
  {\bibinfo {volume} {141}},\ \bibinfo {pages} {211 } (\bibinfo {year}
  {1989})}\BibitemShut {NoStop}%
\bibitem [{\citenamefont {{R. Yang et al.}}(2008)}]{Yang_PRE2008}%
  \BibitemOpen
  \bibfield  {author} {\bibinfo {author} {\bibnamefont {{R. Yang et al.}}},\
  }\href {http://pre.aps.org/abstract/PRE/v78/i6/e066109} {\bibfield  {journal}
  {\bibinfo  {journal} {Phys. Rev. E}\ }\textbf {\bibinfo {volume} {78}},\
  \bibinfo {pages} {066109} (\bibinfo {year} {2008})}\BibitemShut {NoStop}%
\end{thebibliography}

%

\end{document}